\documentclass[journal]{IEEEtran}

\usepackage{amsmath, amssymb, bm, cite, epsfig, psfrag, mathtools}
\usepackage{epstopdf}
\usepackage{graphicx}
\usepackage{algorithm,algpseudocode}
\usepackage{array}
\usepackage{bbm}
\usepackage{multirow}
\usepackage[usenames,dvipsnames]{xcolor}

\usepackage{subfigure}
\usepackage{etoolbox}
\usepackage{pbox}
\usepackage{threeparttable}

\newtoggle{onecol}
\togglefalse{onecol} 
\graphicspath{{figures/}}

\def\beq{\begin{equation}}
\def\eeq{\end{equation}}
\def\beqa{\begin{eqnarray}}
\def\eeqa{\end{eqnarray}}
\def\beqan{\begin{eqnarray*}}
\def\eeqan{\end{eqnarray*}}

\def\Exp{\mathbb{E}}

\def\tm1{t\! - \! 1}
\def\tp1{t\! + \! 1}

\begin{document}

\bibliographystyle{IEEEtran}

\title{Millimeter Wave Cellular Wireless Networks:
Potentials and Challenges}

\author{
    Sundeep Rangan,~\IEEEmembership{Senior Member,~IEEE},
    Theodore S. Rappaport,~\IEEEmembership{Fellow,~IEEE},
    Elza Erkip,~\IEEEmembership{Fellow,~IEEE}
    \thanks{This material is based upon work supported by the National Science
    Foundation under Grants No. 1116589 and 1237821 as well as generous support
    from Samsung, Nokia Siemens Networks, Intel, Qualcomm and InterDigital Communications.}
    \thanks{S. Rangan (email: srangan@nyu.edu), T. Rappaport (email: tsr@nyu.edu) and
            E. Erkip (email: elza@poly.edu) are with the NYU WIRELESS Center,
            Polytechnic School of Engineering, New York University, Brooklyn, NY.}
}

\maketitle

\begin{abstract}
Millimeter wave (mmW) frequencies between 30 and 300~GHz are a
new frontier for cellular communication that offers the promise of
orders of magnitude greater bandwidths combined
with further gains via beamforming and spatial multiplexing
from multi-element antenna arrays.
This paper surveys measurements and capacity studies to
assess this technology with a focus on small cell deployments
in urban environments.
The  conclusions are extremely encouraging;
measurements in New York City at 28 and 73~GHz demonstrate that,
even in an urban canyon environment,
significant non-line-of-sight (NLOS) outdoor, street-level
coverage is possible up to approximately 200~m from a
potential low power micro- or picocell base station.
In addition, based on statistical channel models from these measurements,
it is shown that mmW systems
can offer more than an order of magnitude increase in capacity
over current state-of-the-art 4G cellular networks at current cell densities.
Cellular systems, however, will need to be significantly redesigned
to fully achieve these gains.
Specifically, the requirement of highly directional and adaptive transmissions,
directional isolation between links and significant possibilities of outage
have strong implications on multiple access,
channel structure, synchronization and receiver design.
To address these challenges,  the paper discusses how various
technologies including adaptive beamforming,
multihop relaying, heterogeneous network architectures and carrier aggregation
can be leveraged in the mmW context.
\end{abstract}

\begin{IEEEkeywords}
millimeter wave radio, 3GPP LTE, cellular systems, wireless propagation,
channel models, 28~GHz, 73~GHz, urban deployments.
\end{IEEEkeywords}

\section{Introduction}
\label{sec:intro}

Demand for cellular data has been growing at a staggering pace,
with conservative estimates ranging from 40\% to 70\% year upon year
increase in traffic \cite{CiscoVNI:latest,EricssonMDT:latest,UMTSForecast}.
This incredible growth implies that within the next decades,
cellular networks may need to deliver as much as
1000 times the capacity relative to current levels.
At the same time, as the benefits of wireless connectivity move beyond smartphones
and tablets, many new devices will require wireless service --
perhaps as many as 50 billion devices will be connected by 2020 in one estimate
\cite{Ericsson-50Billion}.
Meeting this demand will be a formidable task.  Many of the requirements
envisioned for what are now being called
Beyond 4G and 5G cellular systems, such as multi-Gbps
peak throughputs and tens of Mbps cell edge rates \cite{NSN-B4G:12},
are already daunting.

\begin{figure}
\begin{center}
    \includegraphics[width=3in]{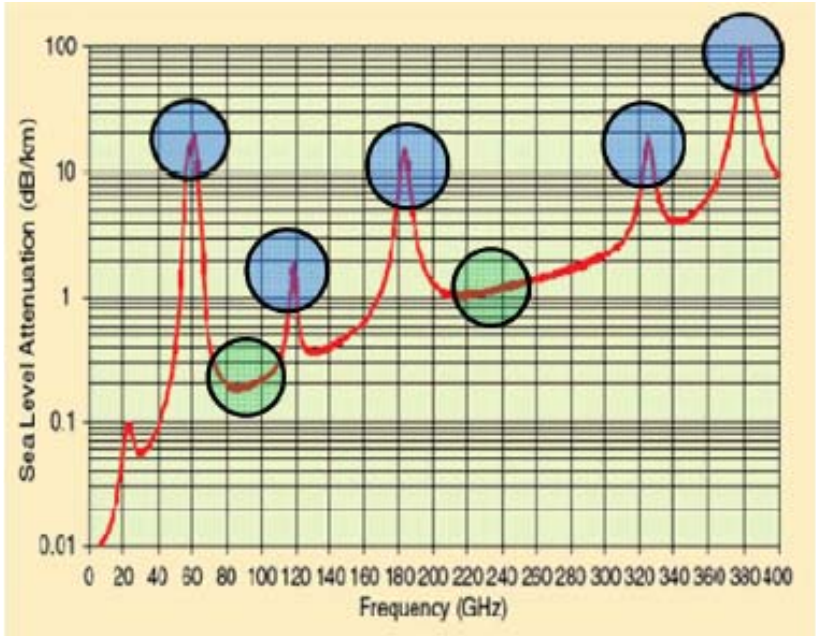}
\end{center}
\caption{\textbf{Millimeter wave (mmW) bands} between 30 and 300 GHz
offer more than 200 times the spectrum than current cellular allocations,
with ample regions with sufficiently low attenuation for small outdoor cells.
In bands with the green bubbles, the
oxygen attenuation is only a fraction of a dB greater than free space over distances of several hundred meters.
Figure from \cite{Ted:60Gstate11}.
}
\label{fig:mmWaveSpec}
\vspace{-0.1in}
\end{figure}

To address this challenge, there has been growing interest in
cellular systems for the so-called \emph{millimeter-wave} (mmW) bands,
between 30 and 300 GHz\footnote{While the mmW spectrum is defined
as the band between 30-300 GHz, industry has loosely considered mmW to be any
frequency above 10~GHz}, where the available bandwidths are much wider than today's
cellular networks~\cite{KhanPi:11,KhanPi:11-CommMag,Ted:60Gstate11,PietBRPC:12}.
The available spectrum at these higher frequencies can be easily 200 times
greater than all cellular allocations today that are largely constrained
to the
prime RF real estate under 3~GHz~\cite{Ted:60Gstate11,KhanPi:11-CommMag}
(See Fig.~\ref{fig:mmWaveSpec}).
Moreover, the very small wavelengths of mmW signals combined with advances
in low-power CMOS RF
circuits enable large numbers ($\geq$ 32 elements) of  miniaturized antennas
to be placed in small dimensions.  These multiple antenna systems
can be used to form very high gain, electrically steerable arrays,
fabricated at the base station, in the skin of a cellphone,
or even within a chip \cite{Doan:04,Doan:05,ZhaLiu:09,gutierrez2009chip,Nsenga:10,Ted:60Gstate11,Rajagopal:mmWMobile,Huang:2008:MWA:1524107,Rusek:13}.
As described in Section~\ref{sec:prior}, these advances will accelerate with the
recent commercialization of 60~GHz WiFi products.
This tremendous potential has led to considerable recent interest in mmW cellular
both in industry
\cite{KhanPi:11,KhanPi:11-CommMag,PietBRPC:12,hur2013millimeter,Samsung5G:13} and academia
\cite{fettweis2005wigwam,laskar2007next,AkoumAyaHeath:12,ZhangMadhow1,ZhangMadhow2,ZhangMadhow3,rappaportmillimeter},
with a growing belief that mmW bands will play a significant role in Beyond 4G and 5G cellular systems
\cite{BocHLMP:14}.

Despite this activity, this interest in mmW is still very recent and the use of
millimeter wave bands remain a largely unexplored frontier for cellular communication.
While every other aspect of cellular mobile technology -- including
processing power, memory, digital communications methods and networking --
have seen tremendous progress since digital cellular systems began some
25 years ago, the carrier frequencies
of those systems remain largely the same.  With today's severe shortage of spectrum, the time is thus ripe
to consider unleashing the capacity in these new bands.

However, the development of cellular networks in the mmW bands
faces significant technical obstacles, and the feasibility of mmW cellular
communication requires careful assessment.
As we will see below,  while the increase in omnidirectional
path loss due to the higher frequencies of mmW transmissions
can be completely compensated through suitable beamforming and directional transmissions, mmW signals
can be severely vulnerable to shadowing, resulting in outages and
intermittent channel quality.
Device power consumption to support large numbers of antennas
with very wide bandwidths is also a key challenge.

The broad purpose of this paper is to survey recent results
to understand the how significant these challenges are,
provide a realistic assessment of how mmW systems can be viable,
and quantify the potential gain they can provide.  We also use the insights
from this evaluation to offer guidance on the research directions needed
for the realization of next-generation cellular systems in the mmW space.

Since the most significant obstacle to mmW cellular is signal
range for non-line-of-sight (NLOS), longer distance links,
a large focus on this paper is in outdoor channel measurements studies.
In particular, we survey our own measurements
\cite{ben2011millimeter,Rappaport:12-28G,Rappaport:28NYCPenetrationLoss,Samimi:AoAD,Nie72G-PIMRC:13,Rappaport:13-BBmmW,rappaportmillimeter}
made in New York City (NYC) in both the 28 and 73~GHz bands
and the statistical models for the channels developed in \cite{AkLiuRanRapEr:13-arxiv}.
NYC provides an excellent test case for mmW propagation studies, since
it is representative of a dense, urban outdoor environment
where mmW system will likely be initially targeted due to the high user
density, small cell radii (typically 100 to 200~m) and lower mobility.
At the same time, NYC is a particularly challenging setting for
mmW propagation since the urban canyon topology
results in a frequent lack of LOS connectivity, severe shadowing
as well as limitations on the height and placement of cells.

As we describe below, our survey of these channel propagation studies shows that,
even in a dense, urban NLOS environment, significant
signal strength can be detected 100~m to 200~m from a base station with
less than 1 Watt of transmit power.  Such distances are comparable to
the cell radii in current urban UHF/microwave cells
and thus we conclude that mmW systems would not necessarily require
greater density for such use cases.
In fact, using a recent capacity analysis of ours in
\cite{AkLiuRanRapEr:13-arxiv} that was based on the NYC experimental data,
we show that mmW cellular systems can offer at least an
order of magnitude increase in capacity relative
to current state-of-the-art 4G networks with comparable cell density.
For example, it is shown that a hypothetical
1~GHz bandwidth time-division duplex (TDD) mmW system could easily provide a 20-fold increase
in average cell throughput in comparison to a 20+20~MHz Long-Term Evolution (LTE) system.
In cellular systems, where even small increases in capacity can be significant,
these gains are truly remarkable.

We also show that
the design of a cellular system based in the mmW range
will need significant changes, more than just simply scaling the carrier frequency
to reach their full potential.
Most significantly, communication will depend extensively on adaptive
beamforming at a scale that far exceeds current cellular systems.
We show that this reliance on highly directional transmissions
has significant implications for
cell search, broadcast signaling, random access and
intermittent communication.  In addition, due to the particular front-end
requirements in the mmW range, support of highly directional
communications also has implications for multiple access and support of small
packet communications.

A related consequence of highly directional transmissions is that
the links become directionally isolated, with interference playing a
much smaller role that in current small cell networks.  One result is that
many of the technologies introduced in the last decade for interference mitigation,
such as coordinated multi-point, intercellular interference
coordination and interference alignment may have limited gains in
mmW systems.
On the other hand, despite rich multipath and scattering,
signal outage may be a larger bottleneck in delivering
uniform capacity, and we discuss various alternate technologies including
multihop relaying, carrier aggregation and heterogeneous networking
to address these issues.

\section{Millimeter Wave Cellular Networks}
    \label{sec:background}

\subsection{The Path to Millimeter Wave Cellular} \label{sec:prior}

For this paper, mmW signals will refer to
wavelengths from 1 to 10~mm, corresponding to
frequencies approximately in the range of 30 to 300~GHz.
Wireless communications in these mmW bands is not new.
Indeed, the first millimeter communications were demonstrated by Jagadis Bose
more than 100 years ago \cite{Bose:27}.
Currently, mmW bands are widely-used for satellite communications
\cite{Roddy:06} and cellular backhaul
\cite{EricssonBackhaul:13,NGNM-Backhaul:12-short,ECC-Backhaul:12}.
More recently, mmW transmissions have been used for
very high throughput wireless LANs and personal area networks systems
 \cite{Ted:60Gstate11,PerCPY:10,VauNic:10,Daniels:10,Baykas-WPAN:11}
in the newly unlicensed 60~GHz bands.
While these systems offer rates in excess of 1~Gbps, the links
are typically for short-range or point-to-point line-of-sight (LOS) settings.

The application of mmW bands for longer range, non-line-of-sight (NLOS)
cellular scenarios is a new frontier
and the feasibility of such systems has been the subject of considerable debate.
While mmW spectrum offers vastly greater bandwidths than current cellular
allocations,
there is a fear that the propagation of mmW signals is much less favorable.
As we will see below, mmW signals suffer from severe shadowing,
intermittent connectivity and will have higher Doppler spreads.
Given these limitations,  there has been considerable skepticism that
mmW bands would be viable for cellular systems
that require reliable communication across longer range and NLOS
paths \cite{Daniels:10,rappaportmillimeter}.

Two recent trends have encouraged a reconsideration of the viability of mmWave cellular.
First, advances in CMOS RF and digital processing have enabled low-cost
mmW chips suitable for commercial mobiles devices
 \cite{Doan:04,Ted:60Gstate11,Rappaport:13-BBmmW}.
Significant progress has been made in particular in power amplifiers and
free-space adaptive array combining, and
these technologies are likely to advance further with the growth of 60~GHz wireless LAN
and PAN systems \cite{Ted:60Gstate11,Daniels:10,PerCPY:10,VauNic:10,Baykas-WPAN:11}.
In addition, due to the very small wavelengths,
large arrays can now be fabricated in a small area
of less than one or two cm$^2$.
To provide path diversity from blockage by human obstructions (such
as a hand holding a part of the device, or the body blocking the path to the cell),
several arrays may be located throughout a mobile device.

Second, cellular networks have been evolving towards smaller radii,
particularly with support for
pico- and femtocell heterogeneous networks in the latest cellular standards
~\cite{Ortiz:08,ChaAndG:08,YehTLK:08,FemtoForum:10,AndrewsCDRC:12}.
In many dense urban areas, cell sizes are now often less than 100~m to 200~m
in radius, possibly within the range
of mmW signals based on our measurements (see Section~\ref{sec:chanMeas}).

In the absence of new spectrum, increasing
capacity of current networks will require even greater ``densification" of cells.
While greater densification is likely to play a central role
for cellular evolution \cite{FemtoForum:10,AndrewsCDRC:12,Qualcomm-SmallCell},
building networks beyond current
densities may not be cost effective in many settings due to
expenses in site acquisition, rollout and delivering quality
backhaul.  Indeed, backhaul already represents 30 to 50\% of the operating
costs by some estimates  \cite{financial_picos,senza_backhaul}
and that share will only grow as other parts of the network infrastructure
decrease in price
\cite{backhaul_blog_1,backhaul_blog_2,financial_picos}.
In contrast, in very high density deployments, the wide bandwidths of mmW signals
may provide an alternative to cell splitting by significantly increasing the capacity
of individual small cells.
Backhaul may also be provided in the mmW spectrum, further reducing costs.

\begin{figure}
    \centering
    \includegraphics[width=3in]{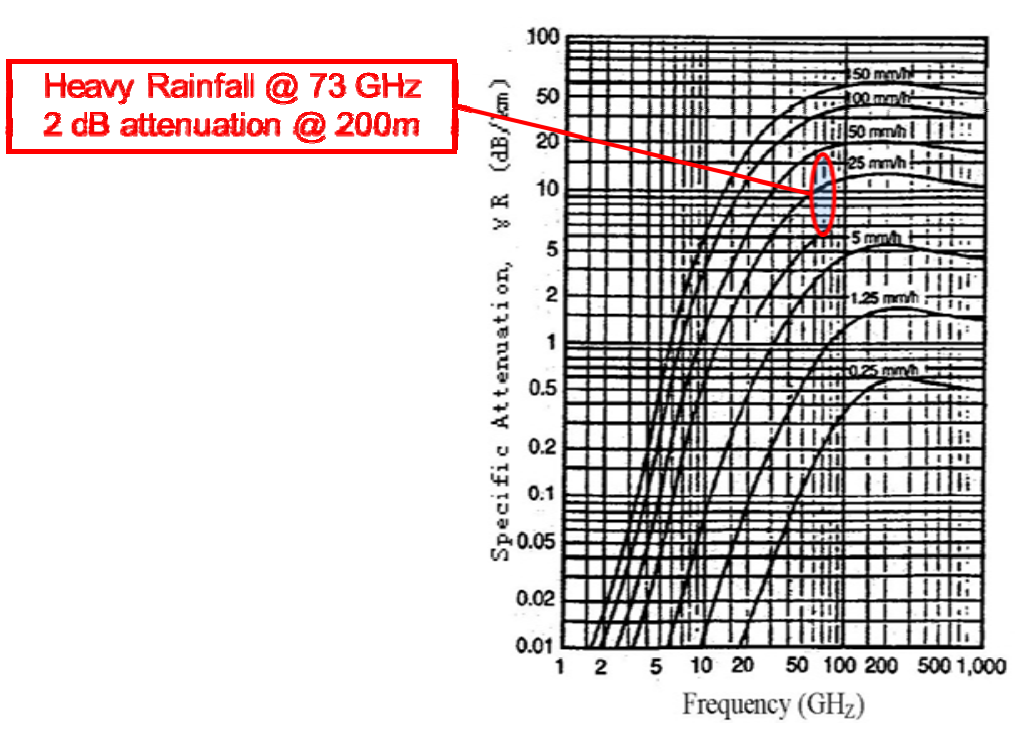}
    \caption{\textbf{Rain fades:}  Even in very heavy rainfall, rain fades
    are typically less than a dB per 100~m meaning they will have minimal impact
    in cellular systems with cell radii less than 200~m.  Figure from \cite{Nie72G-PIMRC:13}.}
    \label{fig:rainFade}
\end{figure}

\subsection{Challenges} \label{sec:challenges}

Despite the potential of mmW cellular systems,
there are a number of key challenges to realizing
the vision of cellular networks in these bands:
\begin{itemize}
\item \emph{Range and directional communication:}
Friis' transmission law \cite{Rappaport:02} states that the free space
omnidirectional path loss grows with the square of the frequency.
However, the smaller wavelength of mmW signals also enables proportionally greater
antenna gain for the same physical antenna size.  Consequently,
the higher frequencies of mmW signals do not in themselves result in any
increased free space propagation loss, provided
the antenna area remains fixed
and suitable directional transmissions are used.  We will confirm this property
from our measurements below -- see also \cite{Sun-Beam:13}.
However, the reliance on highly directional transmissions will necessitate
certain design changes to current cellular systems
that we discuss in Section~\ref{sec:directions}.

\item \emph{Shadowing:}
A more significant concern for range is that
mmW signals are extremely susceptible to shadowing.  For example,
materials such as brick can attenuate signals by as much as 40 to
80~dB\cite{Allen:94,Anderson04,Alejos:08,KhanPi:11-CommMag,Rappaport:28NYCPenetrationLoss}
and the human body itself can result in a 20 to 35~dB loss \cite{LuSCP:12}.
On the other hand, humidity and rain fades -- common problems for long range
mmW backhaul links -- are not an issue in cellular systems
-- See Fig.~\ref{fig:rainFade} and \cite{Ted:60Gstate11,rappaportmillimeter}.
Also, the human body and many outdoor materials
being very reflective, allow
them to be important scatterers for mmW propagation~
\cite{ben2011millimeter,Rappaport:28NYCPenetrationLoss}.

\item \emph{Rapid channel fluctuations and intermittent connectivity:}
For a given mobile velocity, channel  coherence time is
linear in the carrier frequency \cite{Rappaport:02}, meaning that it will be very small
in the mmW range.  For example, the Doppler spread at
60 km/h at 60 GHz is over 3 kHz, hence the channel will change in the order of hundreds of $\mu$s
-- much faster than today's cellular systems.  In addition, high levels of shadowing imply
that that the appearance of obstacles will lead to much more dramatic swings in path loss,
although beamsteering may overcome this \cite{rappaportmillimeter}.
Also, mmW systems will be inherently built of small cells,
 meaning that relative path losses
and cell association also change rapidly.
From a systems perspective, this implies that connectivity will be highly intermittent
and communication will need to be rapidly adaptable.

\item \emph{Multiuser coordination:}  Current applications for mmW transmissions
are for generally for point-to-point links (such as cellular
backhaul \cite{ChiaGB:09}), or LAN and PAN systems
\cite{PerCPY:10,VauNic:10,Daniels:10,Baykas-WPAN:11} with limited number of users
or MAC-layer protocols that prohibit multiple simultaneous transmissions.
However, for high spatial reuse and spectral efficiency,
cellular systems require simultaneous transmissions on multiple interfering
links, and new mechanisms will be needed to coordinate these transmissions
in mmW networks.

\item \emph{Processing power consumption:}
A significant challenge in leveraging the gains of multi-antenna, wide bandwidth
mmW systems  is the power consumption in the A/D conversion.
Power consumption generally scales linearly in the sampling rate
and exponentially in the number of bits per samples~
\cite{Ted:60Gstate11,ChoEtAl_LowPowerAD:94,murdock2014consumption},
making high-resolution quantization at wide bandwidths and large numbers of
antennas prohibitive for low-power, low-cost devices.
For example, scaling power consumption levels of even a state-of-the-art CMOS A/D converter
designs such as \cite{ChenWHL:11,ParkEtAl:11} suggests that
A/Ds at rates of 100~Ms/s at 12b and 16 antennas would require
greater than 250~mW -- a significant drain for current mobile devices.
Also, efficient RF power amplification and combining will be needed for
phased array antennas.

\end{itemize}

\subsection{Deployment Models} \label{sec:network}

\begin{figure}
\begin{center}
\subfigure[Campus setting]{
    \includegraphics[width=1.5in]{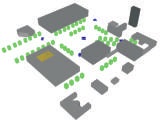} }
\subfigure[Urban picocell]{
    \includegraphics[width=1.5in]{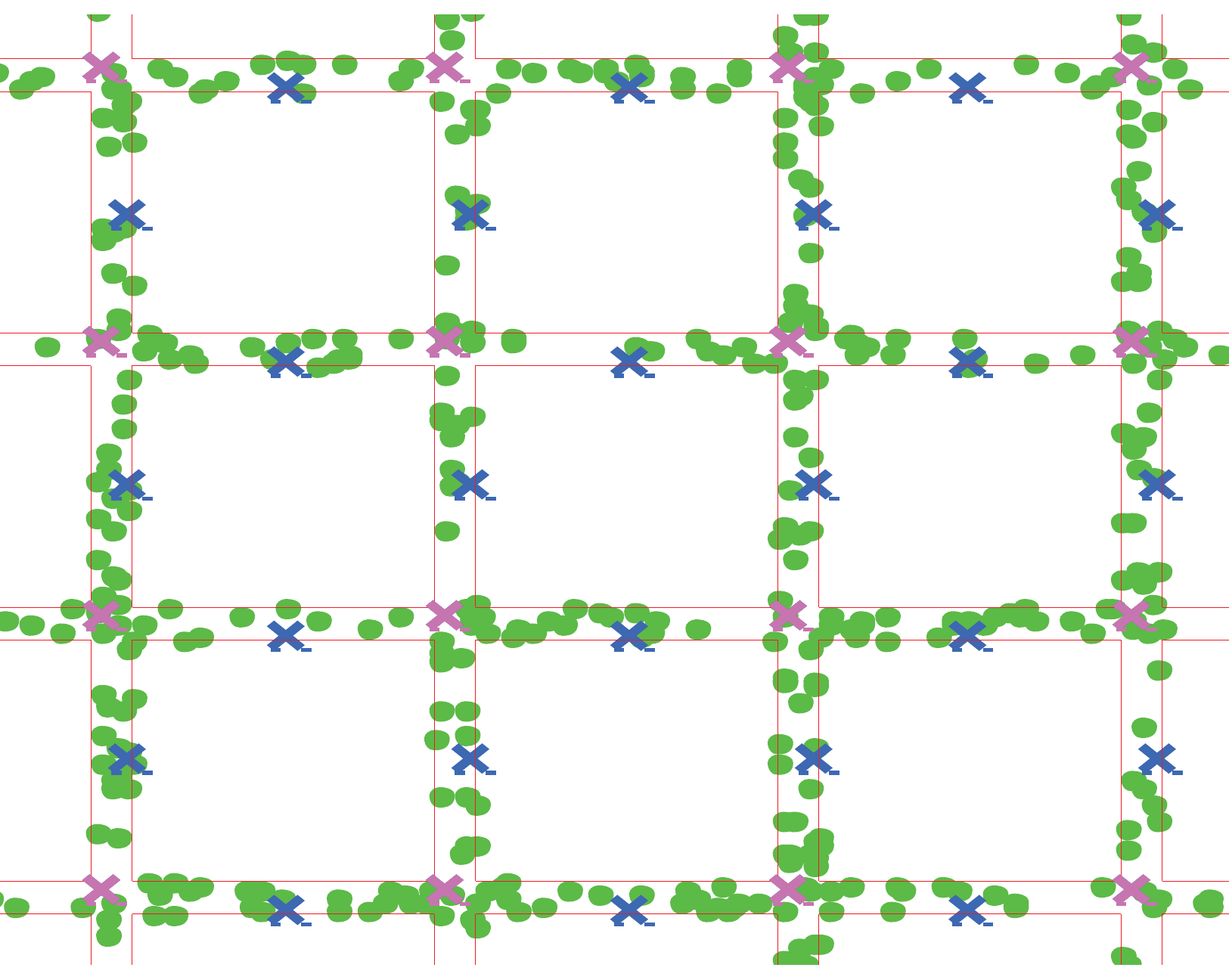} }
\end{center}
\caption{\textbf{Millimeter wave cellular use cases}:
(a) Outdoor coverage in a campus-like environment, as illustrated in
 \cite{abouelseoud2013system}. (b) Urban micro- or picocells
as illustrated in a figure detail from \cite{Ghosh-mmw:2013}
showing mmW access points
(blue and pink crosses) placed placed on every block on an urban grid
to serve mobiles (green circles) on the streets.
}
\label{fig:mmWUseCase}
\end{figure}

Due to the limited range of mmW signals, most of the cellular applications
for mmW systems have focussed on small-cell, outdoor deployments.
For example, a capacity study by Pietraski et.\ al.\ \cite{PietBRPC:12,abouelseoud2013system} considered
deployments in campus and stadium-like settings where the users could
obtain relatively unobstructed connections to the mmW cells
-- See Fig.~\ref{fig:mmWUseCase}(a).

The focus in this paper will be in urban micro- and picocellular deployments
with cell radii in the range of 100~m to 200~m -- similar to current
cell sizes for such deployments.
Coverage in urban environments will encounter NLOS propagation much
more frequently  than outdoor campus or stadium settings,
and is thus significantly more challenging.
To provide dense coverage in such scenarios,
the mmW cells could be deployed, for example, in a picocellular manner
on street furniture such as lampposts or sides of buildings to enable direct coverage
onto the streets with minimal shadowing.
Fig.~\ref{fig:mmWUseCase}(b)
 shows such a picocellular layout for an urban environment
considered in~\cite{Ghosh-mmw:2013} where one to three
mmW access points were placed in each block in a city grid.
Other deployments are also possible.  For example,
cells could be placed similar to current urban microcells
on top of buildings for larger area coverage.

\subsection{Heterogeneous Networking Aspects}
\begin{figure}
\begin{center}
    \includegraphics[width=2.75in]{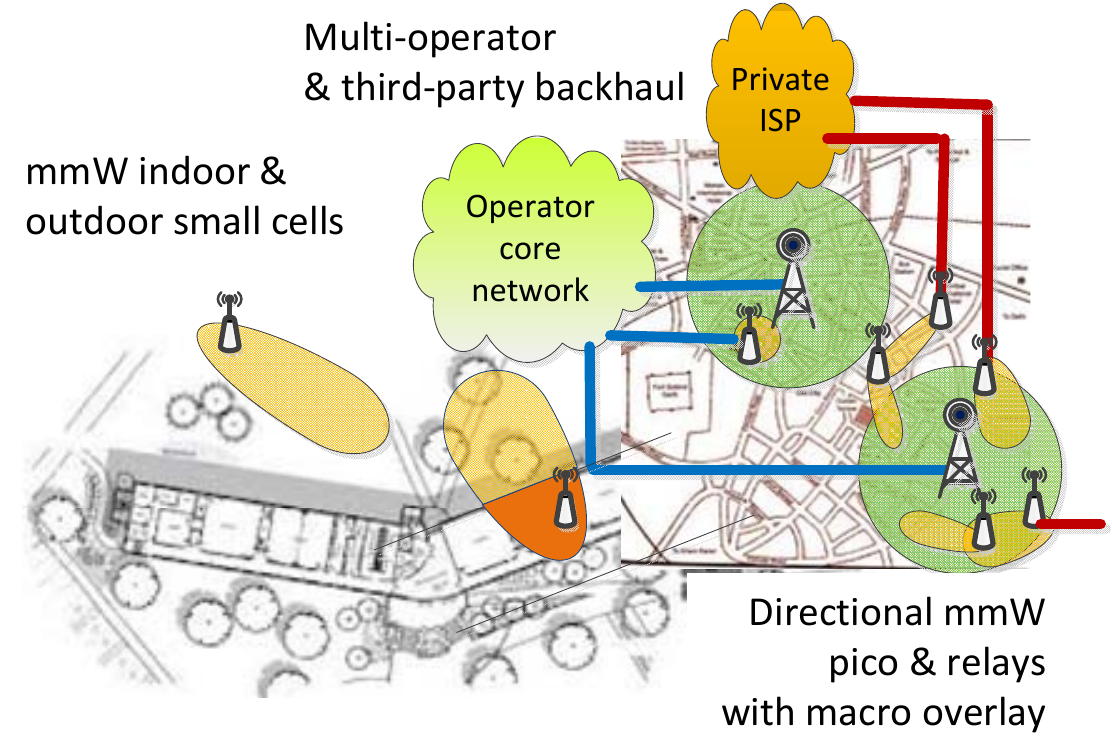}
\end{center}
\caption{Due to the inherent limitations of mmW propagation,
mmW cellular systems will need to co-exist and coordinate with
conventional microwave cells.  Also, to provide indoor coverage and
efficiently use the spectrum, backhaul and spectrum may be shared between
operators and third parties much more significantly than in current
deployments.
}
\label{fig:mmWHetNet}
\vspace{-0.1in}
\end{figure}

Due to the inherent limitations of mmW propagation,
mmW cellular systems cannot alone provide uniform, robust high capacity
across a range of deployments.
Millimeter wave networks will be inherently \emph{heterogeneous} --- See Fig.~\ref{fig:mmWHetNet}.
In fact, it is quite likely that cellular and local area networks
will blur over time.

Heterogeneous networks, or HetNets, have been one of the most active
research areas in cellular standards bodies in the last five years
\cite{QualcommHetNet:11,QCOM-HetNetSurvey:11,ChaAndG:08,AndrewsCDRC:12},
with the main focus being inter-cell interference coordination and load balancing.
However, the introduction of mmW cells into current cellular networks will
create heterogeneity in the network in many more aspects than cell size:
\begin{itemize}
\item \emph{mmW and microwave/UHF}:  Most importantly, since mmW cells
will be inherently limited in range (due to the physical size limitations
of antenna structures and the corresponding gain in a portable device),
they will have to co-exist
with a conventional UHF / microwave cellular overlay  for universal coverage.

\item \emph{Relay vs. wired access points}:  With large numbers of small cells,
it may be impractical or expensive to run fiber connectivity to every cell.
As we will discuss in Section~\ref{sec:relay}, relays
(or, in a simpler from, repeaters) provide an attractive
cost-effective alternative that can build on existing mmW backhaul technology
and exploit the full degrees of freedom in the mmW bands.

\item \emph{Short-range LOS picocells vs.\ NLOS wide-area microcells}:
As described above, there may be significant differences in coverage
between microcells and picocells.  Microcells may offer larger range, but
more diffuse NLOS coverage. In practice, both
cell types will likely need to co-exist~\cite{Rappaport:28NYCPenetrationLoss}.

\item \emph{Ownership}:  A key challenge of mmW is indoor penetration.
Reasonable coverage will require that mmW cells are placed indoors
\cite{Rappaport:28NYCPenetrationLoss,Nie72G-PIMRC:13}.
Analogous to the femtocell concept
\cite{Ortiz:08,ChaAndG:08,YehTLK:08,FemtoForum:10,AndrewsCDRC:12},
and neighborhood small cells \cite{Qualcomm-NSC,FordKR:13-arxiv}, third parties
may be better suited to provide these cells, thereby creating a network
with multiple operators and third-party ownership.

\end{itemize}

Such heterogeneous networks present several design issues,
particularly in cell selection and networking.  We discuss some of these issues in
Section~\ref{sec:hetNet}.

\section{Cellular Channel Measurements} \label{sec:chanMeas}

To assess the feasibility of mmW networks, we begin by surveying
recent channel measurements of mmW signals in urban environments,
particularly our wideband propagation studies in the
28 and 73~GHz bands in New York City.

\subsection{Prior Measurements}

Particularly with the development of 60~GHz LAN and PAN systems, mmW signals have been
extensively characterized in indoor environments
\cite{Ted:60Gstate11,Daniels:10,Zwick05,Giannetti:99,Anderson04,Smulders,Manabe,ben2011millimeter,ted2}.
The propagation of mmW signals in outdoor settings for micro- and
picocellular networks is relatively less understood.

Due to the lack of actual measured channel data,
many earlier studies \cite{KhanPi:11,ZhangMadhow1,AkoumAyaHeath:12,PietBRPC:12}
have relied on either analytic models or commercial ray tracing software
with various reflection assumptions.
These models
generally assume that propagation will be dominated by either LOS links or
links with a few strong specular reflections.  As we will see below,
these models may be inaccurate.

Also, measurements in Local Multipoint Distribution Systems at 28~GHz
-- the prior system most closest to mmW cellular --
have been inconclusive:
For example, a study \cite{ElrefShak:97} found 80\% coverage at ranges
up to 1--2~km, while \cite{SeiArn:95} claimed that LOS connectivity would be
required.  Our own previous
studies at 38~GHz \cite{Rappaport:13-BBmmW,Rappaport38:12,ted:rww12,ted:wcnc12,Rappaport-72GHz:13} found that relatively long-range
links ($> 750$~m) could be established.  However, these measurements
were performed in an outdoor
campus setting with much lower building density and
greater opportunities for LOS connectivity than would be found in a typical
urban deployment.

\subsection{Measurements in New York City} \label{sec:chanMeas28}

\begin{figure}
  \centering
  \includegraphics[width=0.40\textwidth]{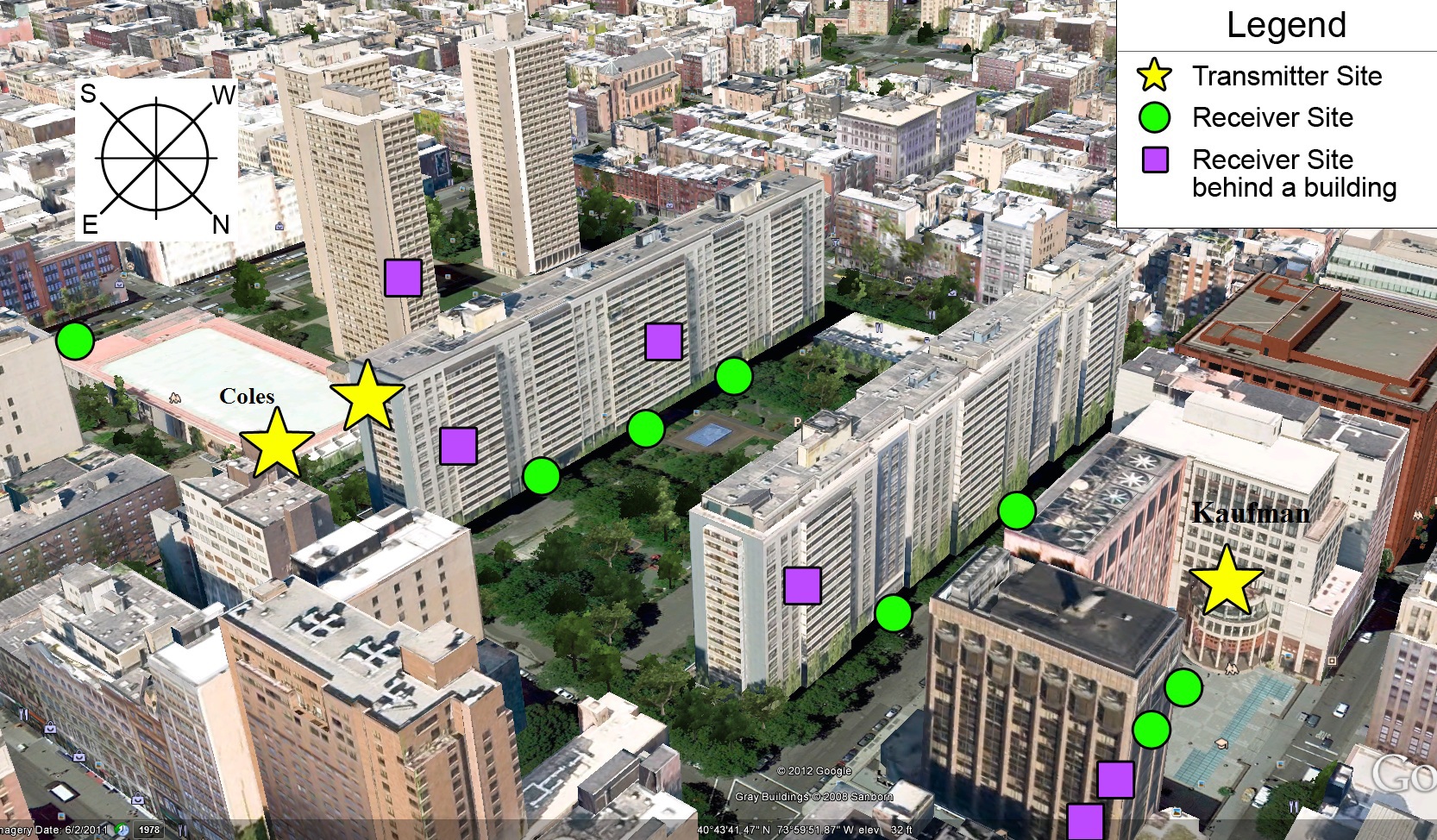}
  \caption{\label{fig:mapMeas} Image from \cite{Rappaport:12-28G}
  showing typical measurement locations in NYC at 28~GHz
  for which the isotropic path  loss models in this paper are derived. Similar locations
  were used for the 73~GHz study.}
\end{figure}

To provide a realistic assessment of mmW propagation in urban environments,
our team conducted extensive measurements of 28 and 73~GHz
channels in New York City.
Details of the measurements  can be found in
\cite{ben2011millimeter,Rappaport:12-28G,Rappaport:28NYCPenetrationLoss,Samimi:AoAD,Nie72G-PIMRC:13,Rappaport:13-BBmmW,rappaportmillimeter,Rappaport-72GHz:13}.

The 28 and 73~GHz bands were selected since they are both likely to be
initial frequencies where mmW cellular systems could operate.
The 28~GHz bands were previously targeted for Local Multipoint Distribution
Systems (LMDS) systems and are now attractive
for initial deployments of mmW cellular given
their relatively lower frequency within the mmW range.
However, as mmW systems become more widely deployed,
these lower frequency mmW bands
will likely become depleted, particularly since they must
compete with existing cellular backhaul systems.
Expansion to the higher bands is thus inevitable.
In contrast, the E-Band frequencies (71-76 GHz and 81-86 GHz)~\cite{wells:09}
have abundant spectrum and
are adaptable for dense deployment, providing a major option
for carrier-class wireless indoor and outdoor transmission, should the lower
frequency become congested.
As shown in Fig.~\ref{fig:mmWaveSpec}, the atmospheric absorption of
E-band is only slightly worse (e.g.\ 1 dB per km)
than today's widely-used lower frequency (UHF/microwave) bands.

\begin{figure}
\begin{center}
    \includegraphics[width=3in]{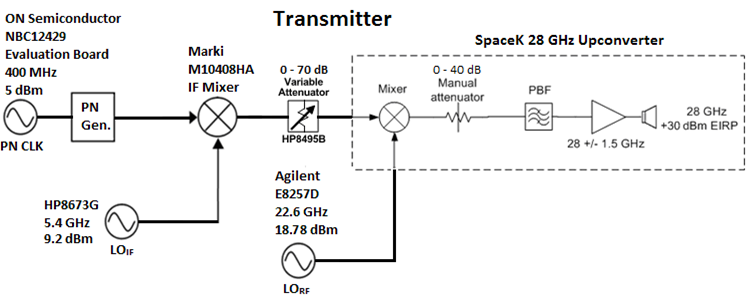}
\end{center}
\caption{28~GHz channel sounder transmitter block diagram with 54.5~dBm EIRP and
800~MHz first null-to-null RF bandwidth for high temporal resolution.
Figure from \cite{Rappaport:12-28G}.
}
\label{fig:ChanSoundTX}
\end{figure}

To measure the channel characteristics in these frequencies,
we emulated microcellular type deployments where
transmitters were placed on rooftops
two to five stories high and measurements were then made at a
number of street level locations up to 500~m from the transmitters
(see Fig.~\ref{fig:mapMeas}).
To characterize both the bulk path loss and spatial structure of the
channels, measurements
were performed with highly directional, rotatable horn antennas (30 dBm RF output,
10 degree beamwidths and 24.5 dBi gain at both TX and RX).
In order to obtain high time resolution, we employed a 400~Mcps channel sounder
(see Fig.~\ref{fig:ChanSoundTX}).
At each transmitter (TX) - receiver (RX) location pair, the angles of the TX and RX antennas
were swept across a range of values to detect discrete clusters of paths
\cite{ben2011millimeter,Rappaport:12-28G,Rappaport:28NYCPenetrationLoss,Samimi:AoAD,Nie72G-PIMRC:13,Rappaport:13-BBmmW,rappaportmillimeter,Rappaport-72GHz:13}.

\subsection{Large-Scale Path Loss Model}

Using the data from
\cite{ben2011millimeter,Rappaport:12-28G,Rappaport:28NYCPenetrationLoss,Samimi:AoAD,Nie72G-PIMRC:13,Rappaport:13-BBmmW,rappaportmillimeter},
detailed statistical models for the channels were developed in our recent work
\cite{AkLiuRanRapEr:13-arxiv},
where we took the directional channel measurements and created narrowband
isotropic (unity gain, omnidirectional)
channel models by adding the powers received over all measurement angles,
and subtracting the 49 dB of original antenna gains used in the measurements.
Here, we summarize some of the main findings from \cite{AkLiuRanRapEr:13-arxiv}
to help understand the potential capacity of mmW systems, and to
identify the key design issues~\cite{Rappaport:13-BBmmW}.

\begin{figure}
    \centering
    \includegraphics[width=3.5in]{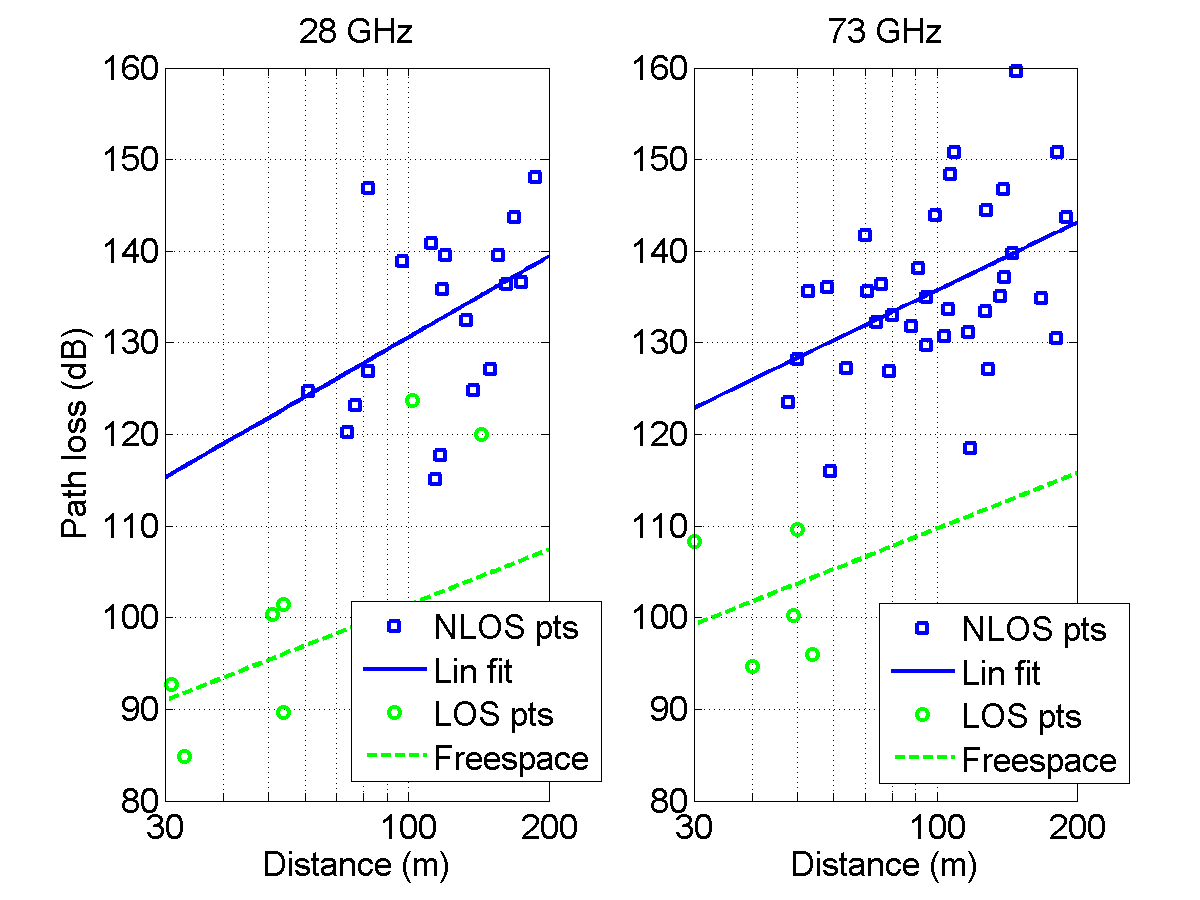}
    \caption{Scatter plot along with a linear fit of the estimated omnidirectional
     path losses as a function of the TX-RX separation for 28 and 73 GHz.
     Figure from \cite{AkLiuRanRapEr:13-arxiv} based on the NYC data in
     \cite{rappaportmillimeter}.}
    \label{fig:distPL}
\end{figure}

First, we summarize the path loss results.
As mentioned above, range is one of the key issues facing mmW systems.
Thus, critical to properly assessing mmW systems, is to first determine how
path loss varies with distance.  Toward this end, Fig.~\ref{fig:distPL}
(taken from \cite{AkLiuRanRapEr:13-arxiv})
shows a scatter plot of the estimated omnidirectional path losses
at different distances from the transmitter.    In both the 28 and 73~GHz measurements,
each point was classified as either being in a NLOS or LOS situation,
based on a manual classification made at the time of the measurements --
see~\cite{ben2011millimeter,Rappaport:12-28G,Rappaport:28NYCPenetrationLoss,Samimi:AoAD,Nie72G-PIMRC:13,Rappaport:13-BBmmW,rappaportmillimeter}.

In standard urban cellular models such as \cite{3GPP36.814}, it is
common to fit the LOS and NLOS path losses separately.
Fig.~\ref{fig:distPL} shows that the LOS path losses roughly follow
the free space propagation based on Friis' Law~\cite{Rappaport:02},
at least for the points with distances $<$ 100~m.
For the NLOS points, \cite{AkLiuRanRapEr:13-arxiv} applied
a standard linear fit of the form
\beq \label{eq:plLin}
    PL(d) \mbox{ [dB]} = \alpha + \beta 10\log_{10}(d) + \xi, \quad \xi \sim {\mathcal N}(0,\sigma^2),
\eeq
where $d$ is the distance in meters, $\alpha$ is the best (MMSE) fit floating intercept
point over the measured distances (30 to 200~m) \cite{Rappaport-72GHz:13},
$\beta$ is the slope of the best fit, and
$\sigma^2$ the lognormal shadowing variance.  The parameter values for
$\alpha$, $\beta$ and $\sigma$ are shown in Table~\ref{tbl:largeScaleParam} along with
other parameters that are discussed below.

Note that a close-in free space reference path loss model with a fixed leverage
point may also be used, which is  equivalent to \eqref{eq:plLin}
with the constraint that $\alpha + \beta 10\log_{10}(d_0)$ has some fixed value $PL(d_0)$
for some close-in free space reference distance $d_0$.
Work in \cite{Rappaport-72GHz:13} shows that this close-in free space model
is less sensitive to perturbations in data and has valuable insights based on
propagation physics for the slope parameter $\beta$ (e.g.\ $\beta=2$ is free space
propagation and $\beta=4$ is the asymptotic path loss exponent for a
two-ray model).  The close-in free space reference model has only a slightly greater
(e.g.\ 0.5 dB larger standard deviation) fitting error.  While
the analysis below will not use this
fixed leverage point, we point this out to caution against
ascribing any physical meaning to the estimated values for $\alpha$ or
$\beta$ in \eqref{eq:plLin} when a floating intercept is used.

\begin{table*}
\caption{Key experimentally-derived model parameters used here and
\cite{AkLiuRanRapEr:13-arxiv} based on the NYC data in  \cite{rappaportmillimeter}. }
\begin{threeparttable}
 \label{tbl:largeScaleParam}

  \begin{tabular}{|>{\raggedright}p{1.5in}|>{\raggedright}p{2in}|
    >{\raggedright}p{1.25in}|>{\raggedright}p{1.25in}|}
	\hline
	\textbf{Variable}  &  \textbf{Model} &
    \multicolumn{2}{c|}{\textbf{Model Parameter Values} }
    \tabularnewline    \cline{3-4}
    & & \textbf{28 GHz} & \textbf{73 GHz}     \tabularnewline  \hline

    Omnidirectional path loss, $PL$    &    $PL = \alpha+10\beta\log_{10}(d) + \xi$,
    $d$ in meters & $\alpha=72.0$, $\beta=2.92$ & $\alpha=86.6$, $\beta=2.45$
        \tabularnewline \hline

	Lognormal shadowing, $\xi$     &  $\xi \sim \mathcal{N}(0,\sigma^2)$ &
     $\sigma = 8.7$ dB & $\sigma=8.0$ dB \tabularnewline \hline


	Number of clusters, $K$ & $K \sim \max\{\mathop{Poisson}(\lambda),1\}$  &
    $\lambda=1.8$ & $\lambda=1.9$ \tabularnewline \hline	

    Cluster power fraction    &    See \eqref{eq:powFrac}.
    & $r_\tau=2.8$, $\zeta=4.0$ & $r_\tau=3.0$, $\zeta=4.0$
        \tabularnewline \hline
    	
    %

	BS cluster rms angular spread & $\sigma$ is exponentially distributed,
 $\Exp(\sigma) = \lambda^{-1}$   &
         \pbox{1in}{ \vspace{2mm} Horiz $\lambda^{-1} = 10.2^\circ$;
         \\ Vert $\lambda^{-1}=0^\circ$ (*) }&
         \pbox{1in}{\vspace{2mm} Horiz $\lambda^{-1} = 10.5^\circ$; \\
          Vert $\lambda^{-1}=0^\circ$ (*) }
         \tabularnewline[3ex] \hline

	UE rms angular spread & $\sigma$ is exponentially distributed,
 $\Exp(\sigma) = \lambda^{-1}$ &
         \pbox{1in}{ \vspace{2mm} Horiz $\lambda^{-1} = 15.5^\circ$;
         \\ Vert $\lambda^{-1}=6.0^\circ$ }&
         \pbox{1in}{Horiz $\lambda^{-1} = 15.4^\circ$; \\
          Vert $\lambda^{-1}=3.5^\circ$ }
         \tabularnewline[3ex] \hline

  \end{tabular}
  \begin{tablenotes}
  \item Note:  The model parameters are derived in
\cite{AkLiuRanRapEr:13-arxiv} based on converting the directional measurements
from the NYC data in  \cite{rappaportmillimeter},
and assuming an isotropic (omnidirectional, unity gain) channel model
with the 49 dB of antenna gains removed from the measurements.
  \item (*) BS downtilt
was fixed at 10 degree for all measurements, resulting in no measurable vertical angular
spread at BS.
  \end{tablenotes}
  \end{threeparttable}
\end{table*}

\begin{figure}
    \centering
    \includegraphics[width=3in]{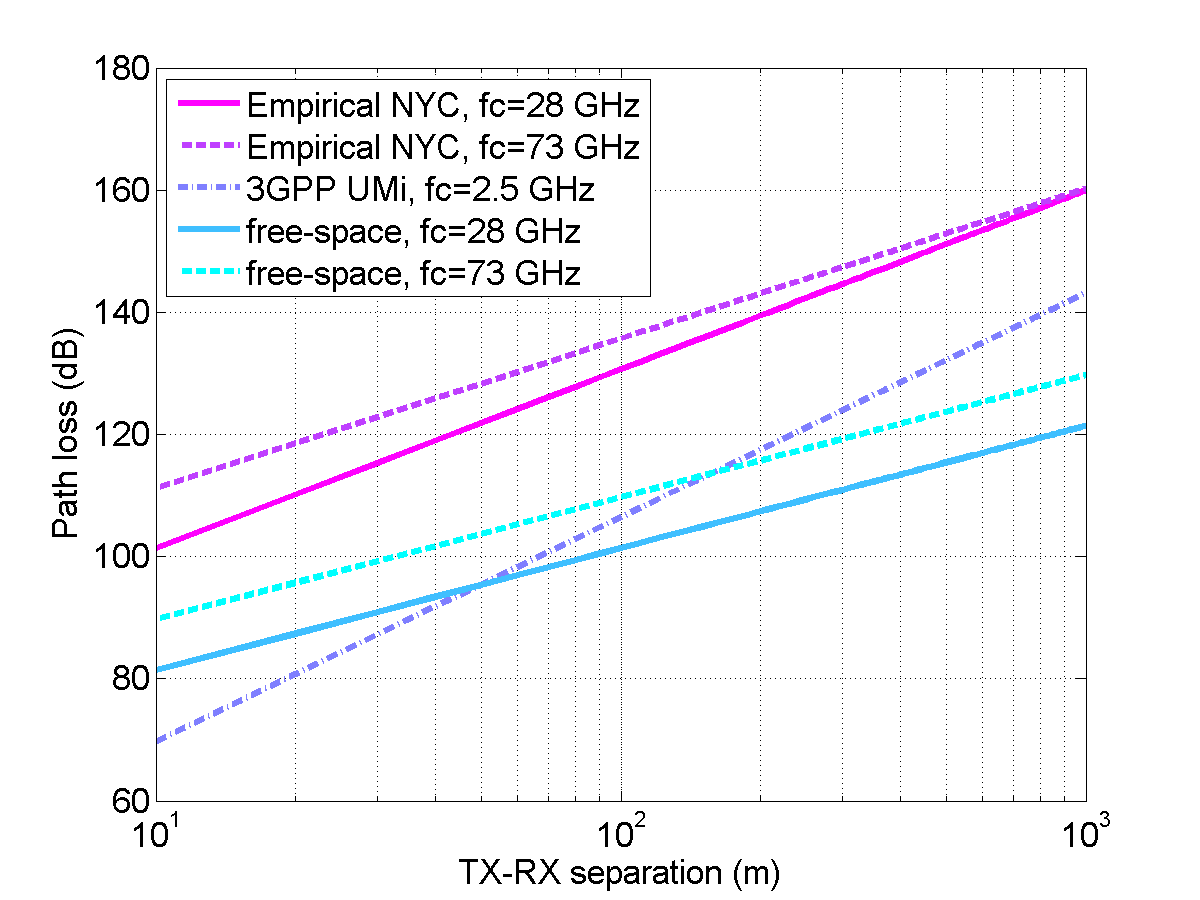}
    \caption{Comparison of distance-based path loss models with unity gain antennas
    from \cite{AkLiuRanRapEr:13-arxiv}.     The curves labeled ``Empirical NYC"
    are the experimentally-derived mmW models based on the NYC data
     \cite{rappaportmillimeter}.
    These are compared to free space propagation for the same frequencies
    and the 3GPP Urban Micro (UMi) model \cite{3GPP36.814} for 2.5~GHz. }
    \label{fig:plComp}
\end{figure}

We can compare the experimentally-derived model \eqref{eq:plLin} for the
mmW frequencies with those used in conventional cellular systems.
To this end, Fig.~\ref{fig:plComp} plots the median
omnidirectional path loss for the following models:
\begin{itemize}
\item \emph{Empirical NYC:}  These curves are the based on the omnidirectional
path loss predicted by our linear model \eqref{eq:plLin} for the mmW channel
with the parameters from Table~\ref{tbl:largeScaleParam},
as derived from the directional measurements in \cite{rappaportmillimeter}.

\item \emph{Free space:}  The theoretical free space path loss is
given by Friis' Law~\cite{Rappaport:02}.
We see, for example, that at $d=$ 100~m, the
free space path loss is approximately 30~dB
less than the omnidirectional propagation model we have
developed here based on the directional measurements in \cite{rappaportmillimeter}.
Thus, many of the works such as
\cite{ZhangMadhow1,AkoumAyaHeath:12,PietBRPC:12}
that assume free space propagation may be somewhat optimistic in
their capacity predictions.
Also, it is interesting to point out that
one of the models assumed in  \cite{KhanPi:11}
(PLF1) is precisely free space propagation + 20 dB -- a correction factor
that is 5 to 10~dB more optimistic than our experimental findings.

\item \emph{3GPP UMi:}  The standard 3GPP urban micro (UMi) path loss
model with
hexagonal deployments~\cite{3GPP36.814} is given by
\beq \label{eq:UMiPL}
    PL(d) \mbox{ [dB]} = 22.7 + 36.7\log_{10}(d) + 26 \log_{10}(f_c),
\eeq
where $d$ is distance in meters and $f_c$ is the carrier frequency in GHz.
Fig.~\ref{fig:plComp} plots this path loss model at $f_c=$ 2.5 GHz.
We see that our propagation models for unity gain antennas
at both 28 and 73~GHz predict
omnidirectional path losses that, for most of the distances,
are approximately 20 to 25~dB higher than the 3GPP UMi model at 2.5~GHz.
However, the wavelengths at 28 and 73~GHz are approximately 10 to 30
times smaller than at 2.5 GHz. Since, for a fixed antenna area,
the beamforming gain grows with $\lambda^{-2}$,
the increase in path loss can be entirely compensated
by applying beamforming at either the transmitter or receiver.
In fact, the path loss can be more than compensated relative to today's cellular systems,
with beamforming applied at both ends.
We conclude that, barring outage events and maintaining the same
physical antenna size, \emph{mmW propagation does not lead to any reduction
in path loss relative to current cellular frequencies and, in fact, may
be improved over today's systems}.
Moreover, further gains may be possible via spatial multiplexing
as we will see below.
\end{itemize}

\subsection{Angular and Delay Spread Characteristics} \label{sec:chanMeasAng}

The channel sounding system, with 10 degree beamwidth rotatable horn antennas and
400~MHz baseband signal bandwidth, enables high resolution time and angular spread
measurements.  One of the key, and surprising, findings of our studies,
was the presence of several distinct clusters of paths
with significant angular and delay spread between the clusters.
This observation provides strong evidence that
-- at least with the microcellular type antennas in an urban canyon-type
 environment -- mmW signals
appear to propagate via several NLOS paths rather than a small number of LOS links.
We note that
these NLOS paths are arriving via reflections and scattering from
different buildings and surfaces
\cite{rappaportmillimeter,ben2011millimeter,Rappaport:12-28G,Rappaport:28NYCPenetrationLoss,Samimi:AoAD,Nie72G-PIMRC:13,Rappaport:13-BBmmW,Rappaport38:12}.

\begin{figure}
\begin{center}
    \includegraphics[width=2.5in]{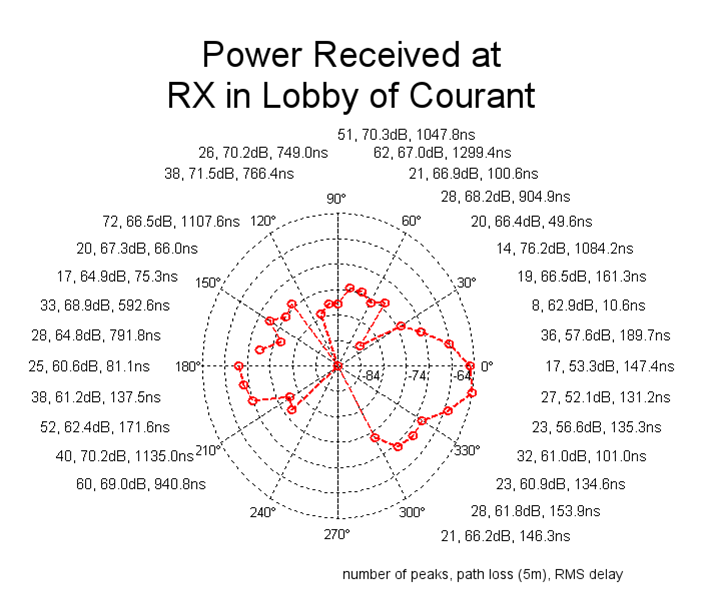} \\
    \vspace{0.5cm}
    \includegraphics[width=2.5in]{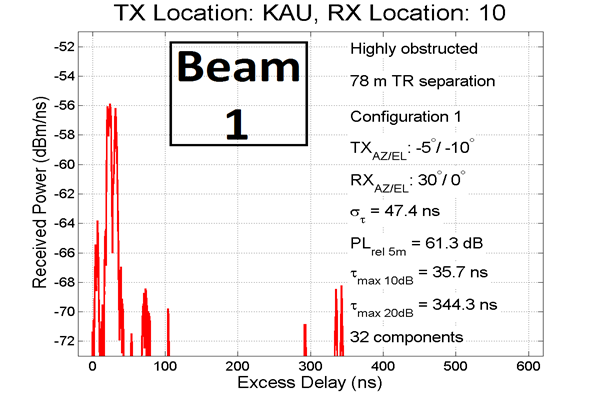}
\end{center}
\caption{Top:  AoA power profile measured in the courtyard
outside a typical building in the 28~GHz measurement campaign.
Bottom:  Power delay profile at a different location.
Figures from \cite{Samimi:AoAD} and \cite{Sun-Beam:13}.
}
\label{fig:AoADelay}
\end{figure}

To illustrate the presence of multiple path clusters,
the top panel of Fig.~\ref{fig:AoADelay} shows the measured
angular-of-arrival (AoA) power profile at a typical location in our 28~GHz measurements.
At this location, we clearly see three angular clusters or ``lobes"
\cite{Samimi:AoAD} -- a common number observed over all locations.
Similarly, the bottom panel shows the power delay profile and we see that
several clusters are apparent.
The presence of discrete clusters, each with relatively narrow angular and delay spread,
will have certain implications for the receiver design that we discuss below in
Section~\ref{sec:compSens}.

The paper \cite{AkLiuRanRapEr:13-arxiv} provides a detailed analysis of the
statistical properties of the paths clusters as based
on the data  \cite{rappaportmillimeter,ben2011millimeter,Rappaport:12-28G,Rappaport:28NYCPenetrationLoss,Samimi:AoAD,Nie72G-PIMRC:13,Rappaport:13-BBmmW}.  Some of the findings are as follows:
\begin{itemize}
\item The number of clusters is well-modeled as a Poisson random variable
with an average of approximately two clusters at each location.
Due to the presence of multiple clusters and angular spread within clusters,
many locations exhibit sufficient spatial diversity to support
potentially two or even three spatial degrees of freedom.
See \cite{AkLiuRanRapEr:13-arxiv} for more details.

\item The angular spread (both between clusters and within clusters)
occurs in the azimuth (horizontal) directions at
both the transmitter and receiver, indicating the presence of local scattering
at both ends.  Some vertical (elevation) angular spread is also observed
at the receivers on the street, potentially from ground reflections.
The root-mean-square (rms) beamspread within each cluster can vary
significantly and is well-modeled via an exponential distribution
with similar parameters as current cellular models such as \cite{3GPP36.814}.

\item The distribution of power amongst the path clusters are well-modeled
via a 3GPP model \cite{3GPP36.814} where the fraction of powers in the $K$
clusters are modeled as random variables $\gamma_1,\ldots,\gamma_K$ with
\beq \label{eq:powFrac}
    \gamma_k = \frac{\gamma_k'}{\sum_{j=1}^K \gamma_j'}, \quad
    \gamma_k' = U_k^{r_\tau-1}10^{-0.1Z_k},
\eeq
where  the first random variable, $U_k\sim U[0,1]$, is uniformly distributed and
accounts for variations in delay
between the clusters (clusters arriving with higher delay tend to have less power),
and the second random variable, $Z_k \sim {\mathcal N}(0,\zeta^2)$, is Gaussian
and accounts for lognormal variations due to difference in shadowing
on different clusters.  The variables $r_\tau$ and $\zeta$ are constants fit
to the observed power fractions.
After fitting the parameters to the data, we found
the main cluster does not have the overwhelming majority of power.  Significant
power is often found in the second or even third strongest clusters,
even considering attenuation due to longer propagation delay \cite{AkLiuRanRapEr:13-arxiv},
again indicating the possibility of spatial multiplexing gains between a single base station
and UE.
\end{itemize}

\subsection{Outage Probability} \label{sec:outage}

Due to the fact that mmW signals cannot penetrate many
outdoor building walls, but are able to reflect and scatter off of them,
signal reception in urban environments relies on
either LOS links or strong reflections and scattering
from building and ground surfaces.
Therefore, a key risk in mmW cellular is outage caused by
shadowing when no reflective or scattering paths can be found~
\cite{Samimi:AoAD,Nie72G-PIMRC:13}.

To assess this outage probability, the study \cite{AkLiuRanRapEr:13-arxiv}
used data from
\cite{rappaportmillimeter,ben2011millimeter,Rappaport:12-28G,Rappaport:28NYCPenetrationLoss,Samimi:AoAD,Nie72G-PIMRC:13,Rappaport:13-BBmmW}
which attempted to find
signals of suitable strength at a number of locations up to 500~m from the transmitter.
Interestingly, the analysis showed that signals were detectable
at all 30 locations in Manhattan within 175~m from the cell.
However, at locations at distances greater than 175~m,
most locations experienced a signal outage.
Since outage is highly environmentally dependent, one cannot generalize
too much from these measurements.  Actual outage may be more significant
if there were more local obstacles, if a human were holding the
receiver in a handheld device
or, of course, if mobiles were indoor.
We discuss some of these potential outage effects below.

\section{Capacity Evaluation and Lessons Learned} \label{sec:capAnal}

Using the experimentally-derived channel models
from the NYC data \cite{rappaportmillimeter},
the paper \cite{AkLiuRanRapEr:13-arxiv}
provided some simple system simulations to assess the potential urban
mmW cellular systems.
We summarize some of the key findings in that work along with other studies
to estimate the possible capacity of mmW systems and
identify the main design issues.

\subsection{System Model}
\begin{table}
\caption{Default network parameters from \cite{AkLiuRanRapEr:13-arxiv}.}
\label{table:para}
\hfill{}
 \begin{tabular}{|>{\raggedright}p{1.25in}|>{\raggedright}p{1.75in}|}
	\hline
	\textbf{Parameter}  &  \textbf{Description} \tabularnewline \hline
	BS layout and sectorization &
        Hexagonally arranged cell sites placed in a 2km x 2km square area
        with three cells per site. \tabularnewline \hline
	UE layout   &   Uniformly dropped in area with average of 10 UEs per
        BS cell (i.e.\ 30 UEs per cell site).
        \tabularnewline \hline
	Inter-site distance (ISD) &    200~m \tabularnewline \hline
    Carrier frequency & 28 and 73 GHz \tabularnewline \hline
    Duplex mode & TDD \tabularnewline \hline
	Transmit power  &  20 dBm (uplink), 30 dBm (downlink) \tabularnewline \hline
	Noise figure  &  5~dB (BS), 7~dB (UE) \tabularnewline \hline
    BS antenna & 8x8 $\lambda/2$ uniform linear array \tabularnewline \hline
    UE antenna & 4x4 $\lambda/2$ uniform linear array for 28~GHz and
          8x8 array for 73~GHz.  \tabularnewline \hline
    Beamforming & Long-term beamforming without single-user or multi-user
    spatial multiplexing \tabularnewline \hline
  \end{tabular}
\hfill{}
\end{table}

Our work here and \cite{AkLiuRanRapEr:13-arxiv} follows a
standard cellular evaluation methodology \cite{3GPP36.814}
where the base stations (BSs) and user equipments (UEs)
are randomly placed according to some statistical model and the performance
metrics were then measured over a number of random realizations of the network.
Since the interest is in small cell networks, we followed a BS and UE distribution
similar to the 3GPP Urban Micro (UMi) model in \cite{3GPP36.814} with some
parameters taken from~\cite{KhanPi:11,KhanPi:11-CommMag}.
The specific parameters are shown in Table~\ref{table:para}.
Observe that we have assumed an inter-site distance (ISD) of 200~m,
corresponding to a cell radius of 100~m.  Also,
the maximum transmit powers of 20~dBm at the UE and
30~dBm at the BS were taken from~\cite{KhanPi:11,KhanPi:11-CommMag}.  These transmit powers
are reasonable since current CMOS RF power amplifiers in the mmW range
exhibit peak efficiencies of at least 8 to 20\% \cite{Ted:60Gstate11,Z3,Z4}.

We considered a network exclusively with mmW cells.  Of course,
in reality, mmW systems will be deployed with an overlay of conventional
larger UHF / microwave cells.  Thus, an actual mmW heterogeneous network
will have a higher capacity, particularly in terms of cell edge rates.
We discuss some of these issues in Section~\ref{sec:hetNet}.

To model the beamforming -- which is essential in mmW systems --  we followed a
conservative model, making
the simplifying assumption that only single stream processing
(i.e.\ no single-user or multi-user spatial multiplexing) was used.
Of course, inter-cell coordinated beamforming and multiple-input multiple-output
(MIMO) spatial multiplexing \cite{ZhangMadhow1,Heath:partialBF} may offer further
 gains, particularly for mobiles close to the cell.  Although these gains
are not considered here, following \cite{Sun-Beam:13}, we considered
multi-beam combining that can capture energy from optimally non-coherently combining
multiple spatial directions
to obtain capacity results here and
in \cite{AkLiuRanRapEr:13-arxiv}.
However, we only considered long-term beamforming \cite{Lozano:07}
to avoid tracking of small-scale fading, which may be slightly challenging at very high
Doppler frequencies (e.g.\ bullet trains) at mmW.

Both downlink and uplink assumed
proportional fair scheduling with full buffer traffic.
In the uplink, it is important to recognize that
different multiple access schemes result in different capacities. If
the BS allows one UE to transmit for a portion of time in the whole band,
the total receive power will be limited to that offered by a single user. If
multiple UEs are allowed to transmit at the same time
time but on different subbands, then the total receive power will be greater,
which is advantageous for users that are not bandwidth-limited. The simulations below thus
assume that subband FDMA is possible.
As we discuss in Section~\ref{sec:multAccess}, this enables much greater capacity
as well as other benefits at the MAC layer.  However, realizing such multiple
access systems presents certain challenges in the baseband front-end which are also
discussed.

\subsection{SINR and Rate Distributions}

\begin{figure}
    \centering
    \includegraphics[width=3in]{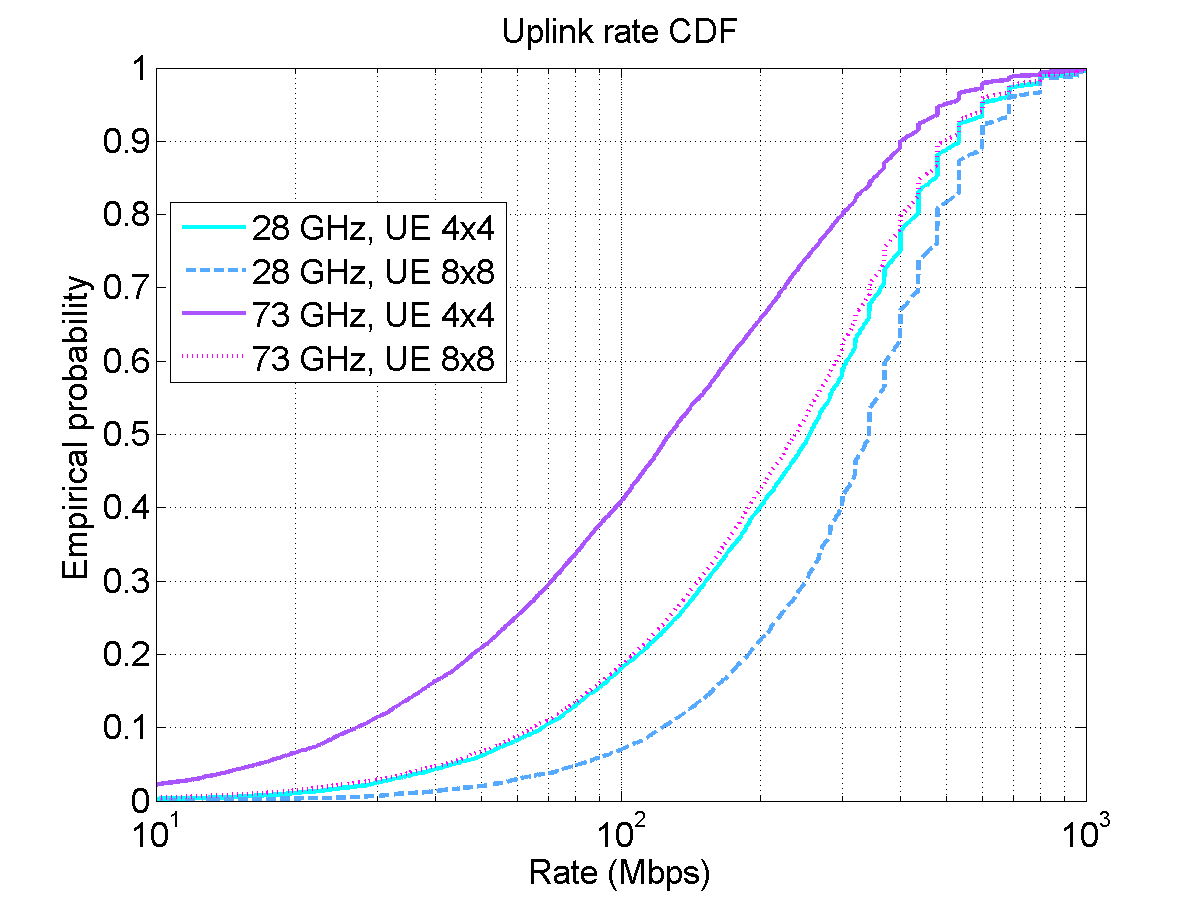}
    \includegraphics[width=3in]{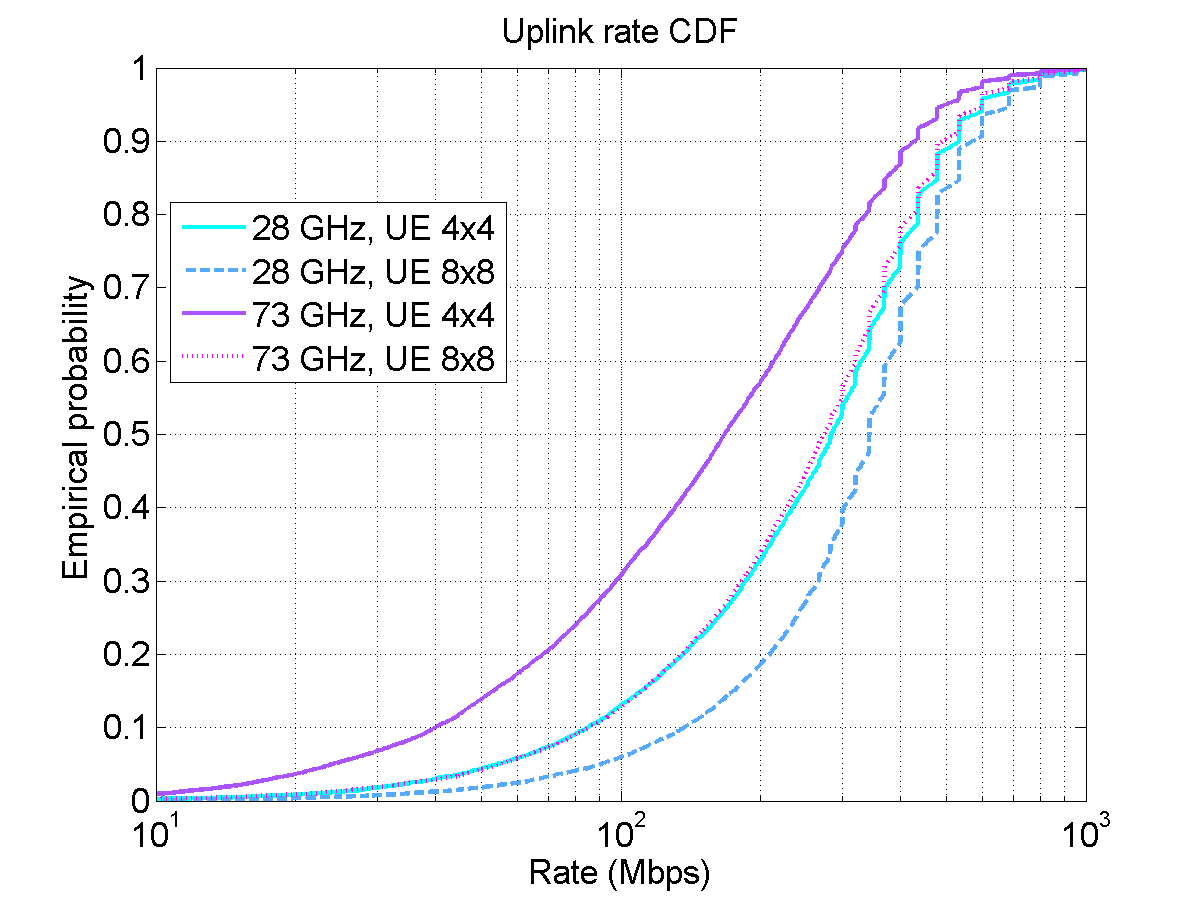}
    \caption{Downlink (top plot) / uplink (bottom plot) rate CDF
    at 28 and 73~GHz with 4x4 and 8x8 antenna arrays at the UE.
    The BS antenna array is held at 8x8.
    Figure from \cite{AkLiuRanRapEr:13-arxiv} based on measurement data in
        \cite{rappaportmillimeter}.}
    \label{fig:rateGeoTxPow}
\end{figure}

We plot SINR and rate distributions in Fig. \ref{fig:sinrGeoTxPow} and Fig. \ref{fig:rateGeoTxPow} respectively. The distributions are plotted for
both 28 and 73~GHz and for 4x4 and 8x8 arrays at the UE.  The BS antenna
array is held at 8x8 for all cases,
although we expect future mmW base stations to have thousands of antenna element
leading much greater gains and directionality.
Some of the key statistics are listed in Table~\ref{table:se}.
More details can be found in \cite{AkLiuRanRapEr:13-arxiv}.

There are two immediate conclusions we can draw from the curves.
First, based on this evaluation, the sheer capacity of a potential
mmW system is enormous.  Cell capacities are often greater than 1~Gbps and
the users with the lowest 5\% cell edge rates experience greater than 10~Mbps.
These rates would likely satisfy
many of the envisioned requirements for Beyond 4G systems such as
\cite{NSN-B4G:12,Ghosh-mmw:2013}.

Second, for the same number of antenna elements, the rates for 73~GHz
are approximately half the rates for the 28~GHz.  However, a 4x4 $\lambda/2$-array
at 28~GHz would take about the same area as an 8x8 $\lambda/2$ array at 73~GHz.
Both would be roughly 1.5$\times$ 1.5 cm$^2$,
which could be easily accommodated in a handheld
mobile device.  In addition, we see that 73~GHz 8x8 rate and SNR distributions
are very close to the 28~GHz 4x4 distributions, which is reasonable since we are
keeping the UE antenna size approximately
constant.  Thus, we can conclude that the
loss from going to the higher frequencies can be made up from larger numbers of antenna
elements without increasing the physical antenna area.

\subsection{Comparison to 4G Capacity}

\begin{figure}
    \centering
    \includegraphics[width=3in]{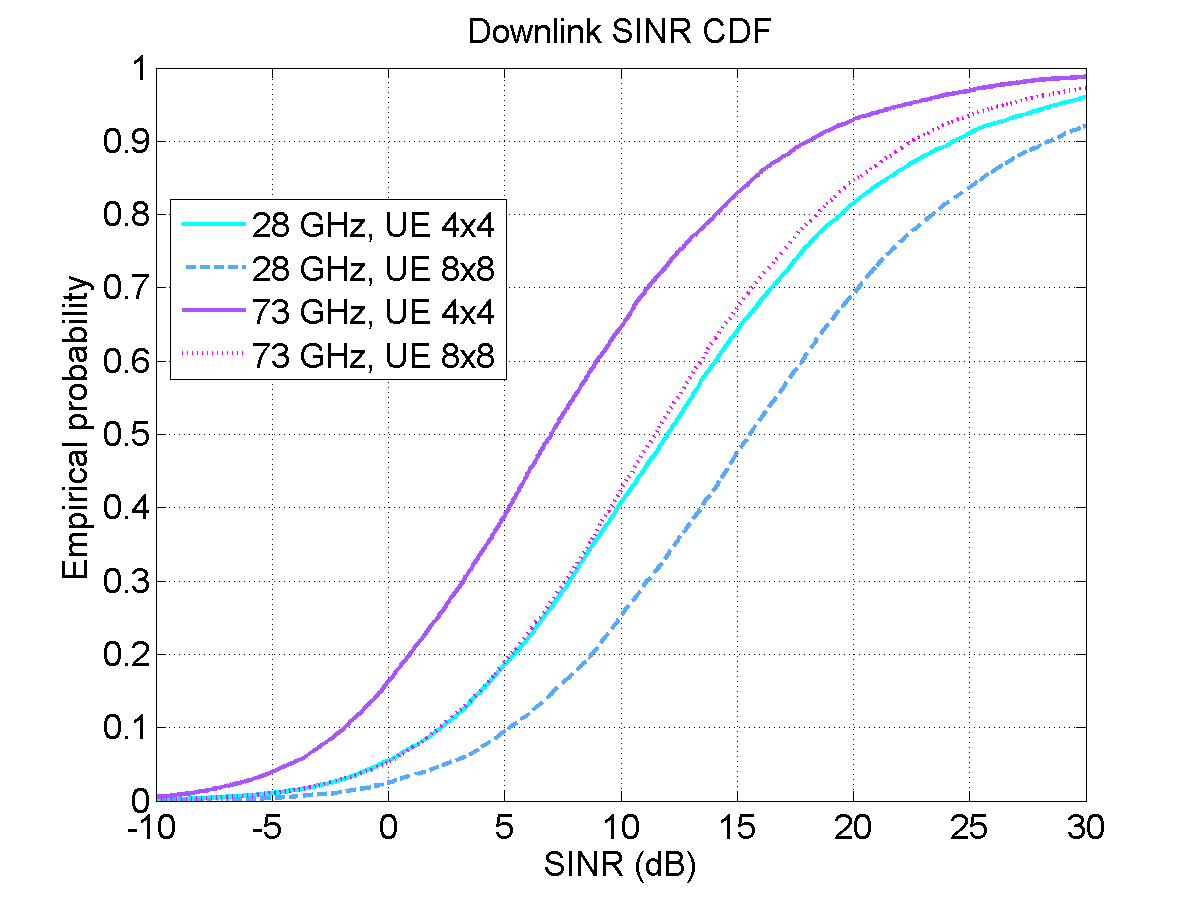}
    \includegraphics[width=3in]{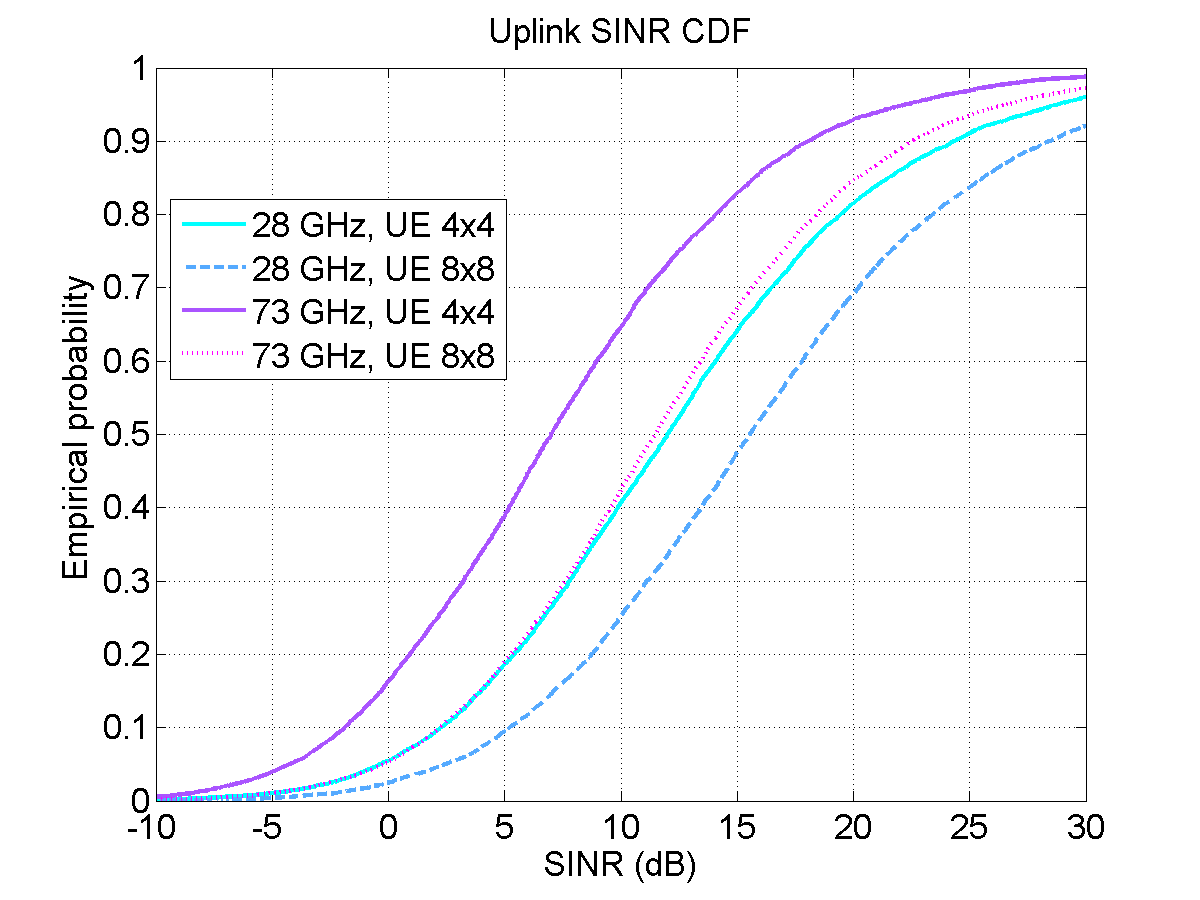}
    \caption{Downlink (top plot) / uplink (bottom plot) SINR CDF
        at 28 and 73~GHz with 4x4 and 8x8 antenna arrays at the UE.
        The BS antenna array is held at 8x8.
        Figure from \cite{AkLiuRanRapEr:13-arxiv} based on measurement data in
        \cite{rappaportmillimeter}.}
    \label{fig:sinrGeoTxPow}
\end{figure}

We can compare the SINR distributions in Fig.~\ref{fig:sinrGeoTxPow}
to those of a traditional cellular network.
Although the SINR distribution for a cellular network in a traditional
UHF or microwave band
is not plotted here, the SINR distributions in Fig.~\ref{fig:sinrGeoTxPow}
are actually slightly better than those found in cellular
evaluation studies \cite{3GPP36.814}.
For example,  in Fig.~\ref{fig:sinrGeoTxPow}, only about 10\% of the mobiles
appear under 0 dB, which is a lower fraction than typical cellular deployments.
We conclude that, although mmW systems have an omnidirectional path loss that
is 20 to 25 dB worse than conventional microwave frequencies, short cell radii combined
with highly directional beams are able to completely compensate for the loss,
and, in fact, improve upon today's systems.


We can also compare the capacity and cell edge rates using the numbers in
Table \ref{table:se}.
The LTE capacity numbers are taken from the average of industry
reported evaluations given in~\cite{3GPP36.814}
-- specifically Table 10.1.1.1-1 for the downlink
and Table 1.1.1.3-1 for the uplink.  The LTE evaluations include
advanced techniques such as spatial division multiple access (SDMA),
although not coordinated multipoint.
For the mmW capacity, we assumed 50-50 uplink-downlink (UL-DL) TDD split and a 20\%
control overhead in both the UL and DL directions.

Under these assumptions, we see from Table~\ref{table:se}
that the spectral efficiency of the mmW system
for either the 28~GHz 4x4 array or 73~GHz 8x8 array is roughly comparable
to state-of-the art LTE
systems\footnote{Note that the spectral efficiency for the mmW
system is quoted including the 20\% overhead, but not the 50\% UL-DL duplexing loss.
Hence, the cell capacity in Table~\ref{table:se} is $C = 0.5\rho W$, where
$\rho$ is the spectral efficiency and $W$ is the baseband bandwidth.}.
Due to its larger bandwidth, we see in Table~\ref{table:se} (cell capacity)
that  mmW systems offer
a significant 20-fold
increase of overall cell capacity.
Moreover, this is a basic mmW system with no spatial multiplexing or other advanced
techniques --  we expect even higher gains when advanced technologies are applied to optimize the mmW system.

While the 5\% cell edge rates are less dramatic, they still offer a 9 to 10
fold increase.
This indicates a significant limitation of mmW systems under NLOS propagation
-- edge of cell users become power-limited and are unable to fully exploit
the increased spectrum.  Thus, other features, such as the use of  repeaters / relays,
will be needed achieve a more uniform performance in mmW systems in these
scenarios.

\begin{table*}
\caption{Conservative mmW and LTE cell capacity/cell edge rate comparison
From \cite{AkLiuRanRapEr:13-arxiv} based on isotropic channel models derived from
   measurement data in \cite{rappaportmillimeter}.}
\label{table:se}
\hfill{}
\begin{threeparttable}
 \begin{tabular}{
     |>{\raggedright}p{0.4in}|>{\raggedright}p{0.5in}|
      >{\raggedright}p{0.4in}|>{\raggedright}p{0.4in}|
      >{\raggedright}p{0.4in}|>{\raggedright}p{0.4in}|
      >{\raggedright}p{0.4in}|>{\raggedright}p{0.4in}|
      >{\raggedright}p{0.4in}|>{\raggedright}p{0.4in}|
      >{\raggedright}p{0.4in}|}
	\hline

	System & \multirow{2}{0.5in}{BW \& Duplex}
    & BS antenna & UE antenna &
    fc (GHz)  &   \multicolumn{2}{c|}{Spec. eff (bps/Hz)} &
      \multicolumn{2}{c|}{Cell capacity (Mbps)}
    &  \multicolumn{2}{c|}{5\% Cell edge rate (Mbps)}
        \tabularnewline \cline{6-11}
    & & & & & DL & UL & DL & UL & DL & UL \tabularnewline \hline
	\multirow{2}{0.5in}{mmW} &  \multirow{2}{0.5in}{1 GHz TDD} &
          8x8 & 4x4 & 28 & 2.25 & 2.38 &  1130 & 1190 & 17.4 & 21.6 \tabularnewline \cline{3-11}
     &  & 8x8 & 8x8 & 28 & 2.83 & 2.84 & 1420 & 1420 & 32.7 & 36.3 \tabularnewline \cline{3-11}
     &  & 8x8 & 4x4 & 73 & 1.45 & 1.65 & 730  & 830  &  6.6 &  9.6 \tabularnewline \cline{3-11}
     &  & 8x8 & 8x8 & 73 & 2.15 & 2.31 & 1080 & 1160 & 16.6 & 22.1
        \tabularnewline \hline

    LTE & 20+20 MHz FDD & \pbox{0.4in}{\vspace{2mm} 2 TX,\\ 4 RX} & 2 & 2.5  & 2.69 & 2.36 &
    53.8 & 47.2 & 1.80 & 1.94 \tabularnewline \hline
  \end{tabular}
  \begin{tablenotes}
   \item Note 1.  Assumes 20\% overhead and 50\% UL-DL duty cycle for the mmW system
   \item Note 2. Long-term, non-coherent beamforming are assumed at both the
   BS and UE in the mmW system.
   However, the mmW results assume no spatial multiplexing gains, whereas the LTE results from
   \cite{3GPP36.814} include spatial multiplexing and beamforming.
    \end{tablenotes}
    \end{threeparttable}
\hfill{}
\end{table*}

\subsection{Interference vs.\ Thermal Noise}

\begin{figure}
    \centering
    \includegraphics[width=3in]{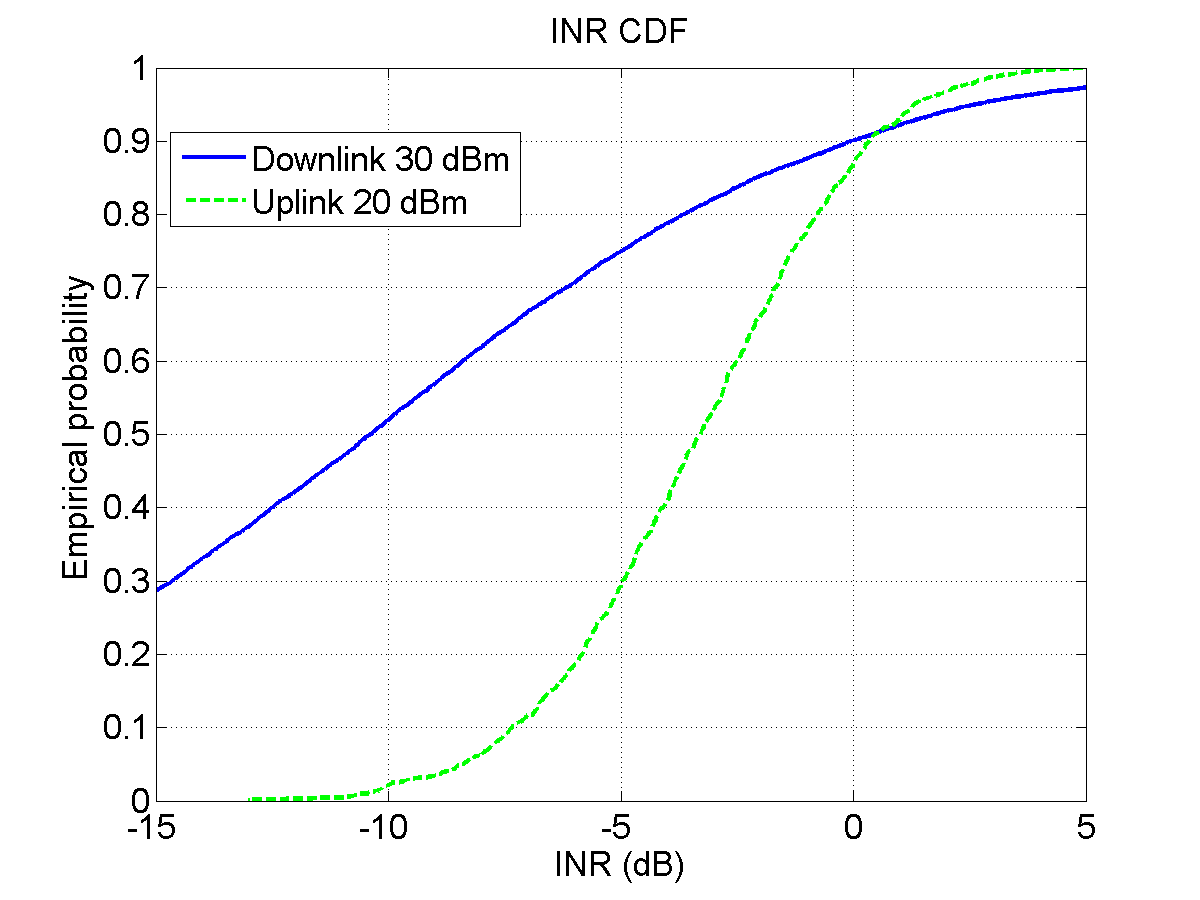}
    \caption{Interference-to-noise ratio in the uplink and downlink
    for the 28~GHz with a 4x4 UE antenna array.}
    \label{fig:inrCDF}
\end{figure}

A hallmark of current small cell systems in urban environments is that they
are overwhelmingly interference-limited, with the rate being limited
by bandwidth, and not power.  Our studies reveal that mmW small cell systems
represent a departure from this model.  For example, Fig.~\ref{fig:inrCDF}
plots the distribution of the interference-to-thermal noise (INR) for
both the uplink and downlink in our simulation of the mmW system
at 28~GHz.  We see that interference is not dominant.  In fact, for the majority
of mobiles, thermal noise is comparable or even larger,
particularly in the downlink.

At the same time, although interference is not dominant, many of the mobiles
are in a bandwidth-limited, rather than power-limited regime.  For example,
Table~\ref{table:se} shows that the average spectral efficiency is
approximately 2.1 to 2.4 bps/Hz in the uplink and downlink for 4x4 28 GHz
or 8x8 73 GHz systems.
We find from Table \ref{table:se} that, if spatial multiplexing is not exploited,
links will be bandwidth-limited and not power-limited, even though interference
is not dominant.
We conclude that, without spatial multiplexing,
mmW systems would represent a new network operating point
not seen in current urban cellular deployments:  large numbers of mobiles would
experience relatively high SINR in directionally isolated links.
In a sense, mmW takes us ``back to the future" when cellular
was first deployed in virgin spectrum.

Of course, without exploiting spatial multiplexing systems would not
benefit from all the degrees of freedom.
We have not yet evaluated single or multi-user MIMO, but such techniques
would lower the SINR per stream for the higher SINR mobiles.  However, the
INR distribution would not significantly change since the total transmit power
would be constant.  Therefore, the links would remain limited by thermal noise
rather than interference.


\subsection{Effects of Outage}

As mentioned above, one of the significant risks of mmW systems
is the presence of outage -- the fact that there is a non-zero probability
that the signal from a given base station can be too weak to be detectable.

To quantify this effect, the paper \cite{AkLiuRanRapEr:13-arxiv}
estimated the capacity under various outage probability models.
The simulations above assumed that at distances greater than a threshold of
$T=175$m, the signal would not be detectable and hence the link
would be in outage.  This assumption was based on the data we observed in
\cite{rappaportmillimeter,ben2011millimeter,Rappaport:12-28G,Rappaport:28NYCPenetrationLoss,Samimi:AoAD,Nie72G-PIMRC:13,Rappaport:13-BBmmW}.
However, as discussed in Section~\ref{sec:outage}, mobiles in
other environments may experience outages closer to the cell,
particularly if there is a lot of ground clutter or the humans themselves
blocking the signal.
To model this scenario, the paper \cite{AkLiuRanRapEr:13-arxiv}
considered a hypothetical outage model, loosely based on
 \cite{3GPP36.814},
 where there was a significant outage probability even close to the cell.
For example, in this model (called a ``soft outage" for reasons explained in
\cite{AkLiuRanRapEr:13-arxiv}), there was approximately a 20\% probability
that a link to a cell would be in outage even when it was only 80~m from the cell.

Interestingly, under this more conservative outage model, the average cell
capacity was not significantly reduced.  However, both the uplink and downlink
5\% cell edge rates fell by a dramatic 50\%.  This reduction shows that mmW systems are robust enough
that mobiles in outage to any one cell will still be able
to establish a connection to another cell.  On the other hand, in environments
where the outages close to the cell are frequent, the gains of mmW systems
will not be nearly as uniform, with cell edge users suffering significantly.

\subsection{Other Studies}
Although our study here and in \cite{AkLiuRanRapEr:13-arxiv}
was the first to use
the experimentally-derived omnidirectional channel models from the
directional data in
\cite{rappaportmillimeter}, the results in
\cite{AkLiuRanRapEr:13-arxiv} roughly
corroborate the findings of very high capacity from mmW systems
predicted in several earlier analyses:
For example, the study in \cite{KhanPi:11} estimated approximately 300~Mbps per cell throughput
in a 500~MHz system.  This capacity corresponds to
a somewhat lower spectral efficiency than what we show here and in
\cite{AkLiuRanRapEr:13-arxiv},
but  the study in \cite{KhanPi:11} assumed only minimal beamforming at the receiver
(either no beamforming or a 2x2 array) and a much larger cell radius of 250~m.

In \cite{PietBRPC:12},
ray tracing software is used to analyze
a mmW campus network and  a median total system capacity
of 32~Gbps with five cell sites, each cell site having four cells, is found.
Since the bandwidth in that study was 2~GHz, the spectral efficiency was
approximately $32/5/4/2=0.8$ bps/Hz/cell.  This number again is lower than our predictions,
but \cite{PietBRPC:12} was limited to QPSK modulation.
Somewhat higher capacity numbers were found in a follow up
study \cite{abouelseoud2013system} in both campus and urban environments.
A later study presented in \cite{Ghosh-mmw:2013} predicted average
spectral efficiencies of almost 1.5 bps/Hz in a 2~GHz system
in an urban grid deployment, a number only slightly lower than our
value of 2.3 to 2.8 bps/Hz.
In all these studies, the cell edge rates
compare similarly to the predicted values in
\cite{AkLiuRanRapEr:13-arxiv}, assuming one normalizes to the
number of users in each cell.

In a different work, \cite{AkoumAyaHeath:12} used a stochastic geometry analysis
and predicted almost 5.4 bps/Hz, which is almost twice our estimated spectral
efficiency.  However, that work assumed that
all links can operate at the Shannon limit
with no maximum spectral efficiency.

This comparison illustrates that, in a number of different scenarios
and analysis methods, the absolute spectral efficiency and
cell edge rate numbers are roughly comparable with estimates here and
in  \cite{AkLiuRanRapEr:13-arxiv} that used experimentally-derived
channel models.
Thus, the broad message remains the same:  under a wide variety of simulation
assumptions, mmW systems can offer orders of magnitude increases in capacity
and cell edge rate over state-of-the-art systems in current cellular bands.

\section{Key Design Issues and Directions for mmW 5G} \label{sec:directions}

The above preliminary results show that while
mmW bands offer tremendous potential for capacity, cellular systems may need
to be significantly redesigned.
In this section, we identify several key design issues that need to be addressed
from a systems perspective if the full gains of mmW cellular systems are to be achieved.

\subsection{Directional Transmissions and Broadcast Signaling}

The most obvious implication of the above results
is that the gains of mmW system depend on highly directional transmissions.
As we discussed above, directionality gains with appropriate beamforming
can completely compensate for, and even further reduce, any increase in the
omnidirectional path loss with frequency.
Indeed, once we account for directional gains enabled by smaller wavelengths,
the path loss, SNR and rate distributions in the mmW range
compare favorably with (and may improve upon)  those in current cellular frequencies.

One particular challenge for relying on
highly directional transmissions in cellular systems
is the design of the synchronization and broadcast signals used in the
initial cell search.
Both base stations and mobiles may need to scan over a range of angles
before these signals can be detected.
This ``spatial searching" may delay base station detection in handovers
--- a point made in a recent paper \cite{LimmWBooster:13}.
Moreover, even after a mobile has detected a base station,
detection of initial random access signals from the mobile may be delayed
since the base station may need to be aligned in the correct direction.

A related issue is supporting intermittent communication
(say through discontinuous reception and transmission (DRX and DTX) modes)
which has been essential in standards such as LTE
for providing low power consumption with ``always on" connectivity~\cite{bontu2009drx}.
In order that either a mobile or base station can quickly begin transmitting,
channel state information in the form of
the spatial directions will need to be maintained at the transmitter.
If cells are small, even the second-order spatial statistics of the channel
may change relatively fast implying some sort of intermittent transmissions
may need to be performed to track the channel state.

\subsection{Multiple Access and Front-End / Baseband Considerations} \label{sec:multAccess}

With small cells, the need for future spectrum / bandwidth flexibility,
support for beamforming and
low cost, TDD (time division duplex) is an attractive duplexing strategy for mmW.
Our analysis in Table~\ref{table:se} assumes TDD for mmW.

\begin{figure}
    \centering
    \includegraphics[width=3in]{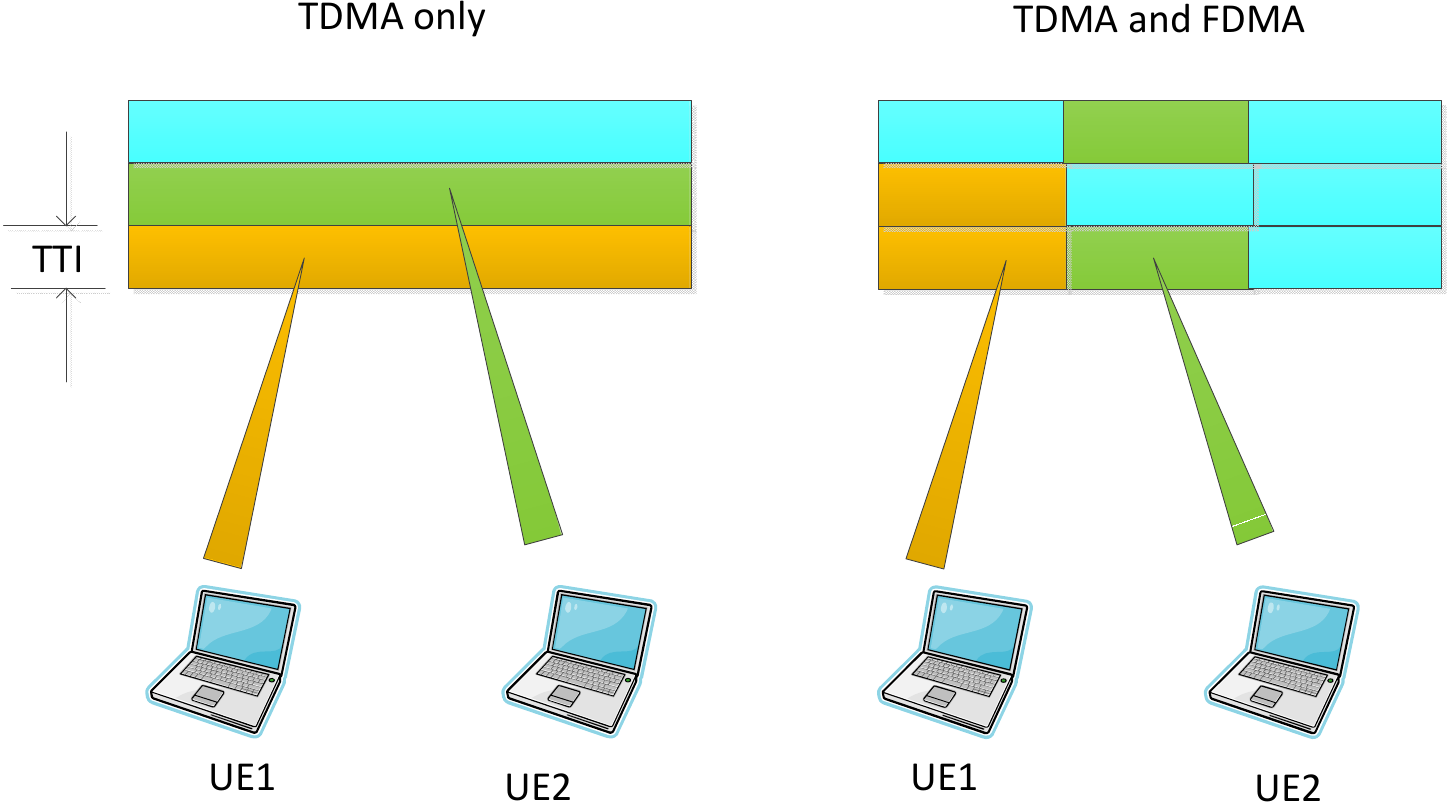}
    \caption{\textbf{Multiple access}:  Enabling FDMA (within a TDD time slot),
    where multiple UEs can be scheduled
    at a time, can offer numerous benefits in mmW systems including improved
    power in the uplink, more efficient transmission of small packets and
    reduced UE power consumption.  A key design issue is how to support FDMA in TDD
    with mmW front-ends that perform beamsteering in analog.}
    \label{fig:FDMA}
\end{figure}

However, closely related to the issue of directional transmissions is how to support
frequency-division multiple access within the TDD time slots.
Current cellular systems use digital processing for MIMO and beamforming.
However, with the large numbers of antennas and wide bandwidths,
it is simply not practical from a power or cost perspective to place
high-resolution, wideband ADCs on each antenna element in the mmW range
\cite{Ted:60Gstate11,KhanPi:11-CommMag,KhanPi:11}.
Most commercial designs have thus assumed phased-array architectures where signals
are combined either in free space or
RF with phase shifters \cite{ParkZim:02,KohReb:07,KohReb:09}
or at  IF \cite{Crane-Patent:88,RamBaRe:98,GuanHaHa:04} prior to the A/D conversion.
A limitation of such architectures is that they
will forgo the support of spatial multiplexing and multi-user
transmissions within the TDD time slots and require time-division
multiple access (TDMA) with only one user within a time slot being scheduled at a time.
In particular, FDMA transmissions within the same time slot
as supported in LTE through resource blocks
will not be possible -- See Fig.~\ref{fig:FDMA}.

\begin{figure}
    \centering
    \includegraphics[width=0.40\textwidth]{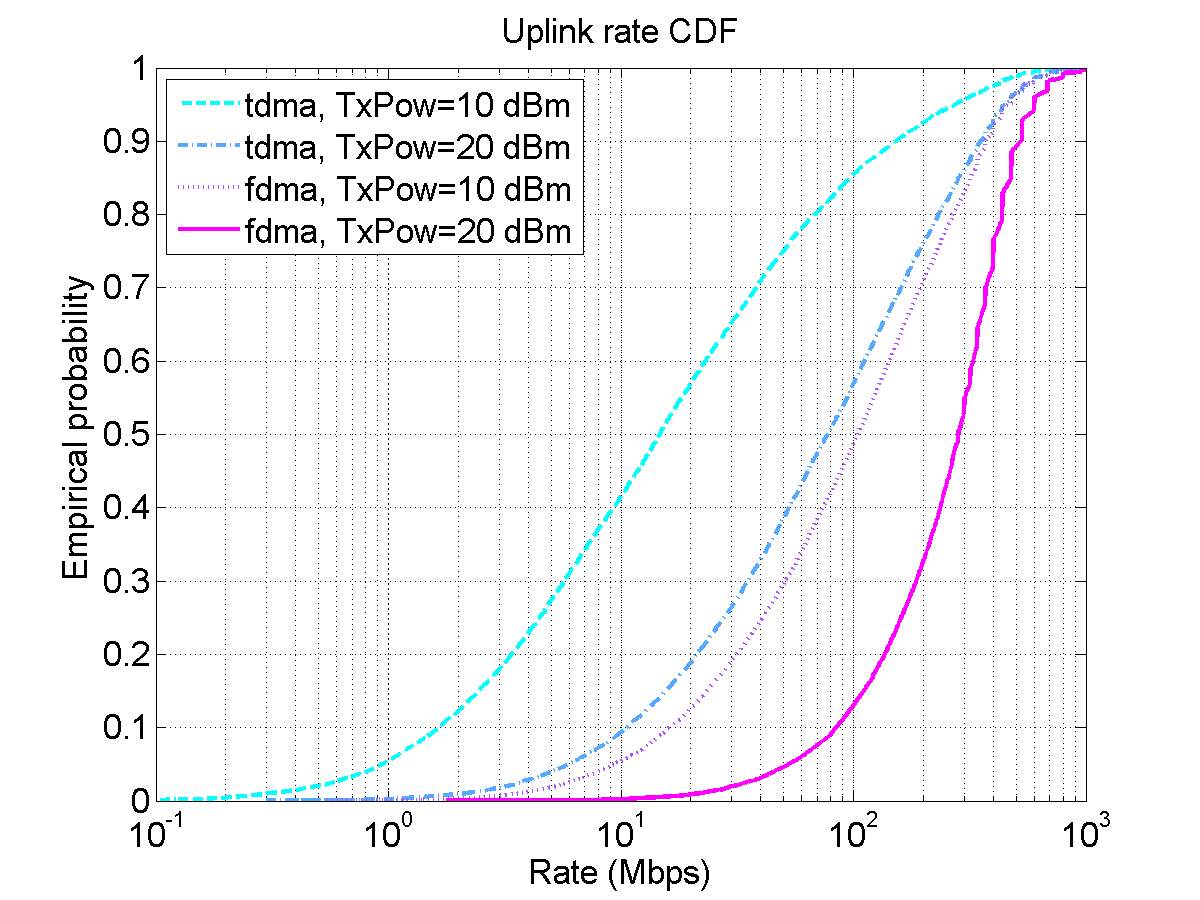}
    \caption{\textbf{Power loss with TDMA only:}  Designs that
    do not enable multiple users to be scheduled at the same time can suffer a significant
    penalty in capacity in the uplink due to loss of power.  Shown here is the rate
    distribution comparing FDMA and TDMA scheduling using beamforming with the 28~GHz isotropic
    channel model. }
    \label{fig:rateGeoMA}
\end{figure}

Enabling granular allocations in frequency is one of the main
hallmarks of LTE, and sacrificing this capability by restricting to TDMA scheduling
will bear significant costs in mmW:
\begin{itemize}
\item \emph{Uplink power:}
Restricting to TDMA scheduling within a TDD time slot implies that
the power of only UE can be received at a time.
Since mobiles at the cell edge may be power-limited,
this reduction of power can significantly reduce capacity.
For example,
according to the uplink rate CDF shown in Fig.~\ref{fig:rateGeoMA}, one can easily see an order of magnitude improvement when multi-user transmission is enabled by FDMA, compared to a baseline TDMA,
both assuming TDD.

\item \emph{Support for small packets:}
Supporting multi-user transmissions will also be essential
to efficiently support messages with small payloads
and is needed for low latency machine-to-machine communications \cite{fettweis2011entering}.
Specifically, when only one UE can transmit
or receive at a time, it must be allocated the entire bandwidth
in a TDD slot, which is extremely
wasteful for small packets.  As an example, in the design of \cite{KhanPi:11-CommMag},
the transmission time interval (TTI) is 125$\mu$s.  Thus, a 1~GHz allocation at this TTI
will have approximately 125,000 degrees of freedom.  Such large transport blocks
would be terribly inefficient, for example, for TCP ACKs as well as other control signaling.

\item \emph{Power consumption:}
From a power consumption perspective, it may be preferable for individual UEs
to only process only a smaller portion (say 100~MHz) of the band
during a time slot.  Such subband allocations
can reduce the power consumption of the baseband processing
which generally scales linearly
in the bandwidth.
\end{itemize}

Thus, a key design issue facing 5G mmW systems is how to support multiple access while
enabling low power consumption, particularly at the UE.
One promising route has been the use of compressed sensing and other
advanced low-bit rate technologies, suggested in \cite{Madhow:ADC}.

In addition, one may consider other SDMA algorithms that optimally exploit
a smaller number of beams.  For example, each UE can still support only one digital stream,
potentially on a subband for low power consumption.  The base station, which
would generally have somewhat higher power capacity, could support a smaller
number, say $K$, beams.  Then, to support $N$ UEs with $K < N$, the base station
can simply select the $K$ beams to span the ``best" $K$-dimensional subspace
to capture the most energy of the $N$ users.

\subsection{Directional Relaying and Dynamic Duplexing} \label{sec:relay}

\begin{figure}
\begin{center}
\includegraphics[width=3in]{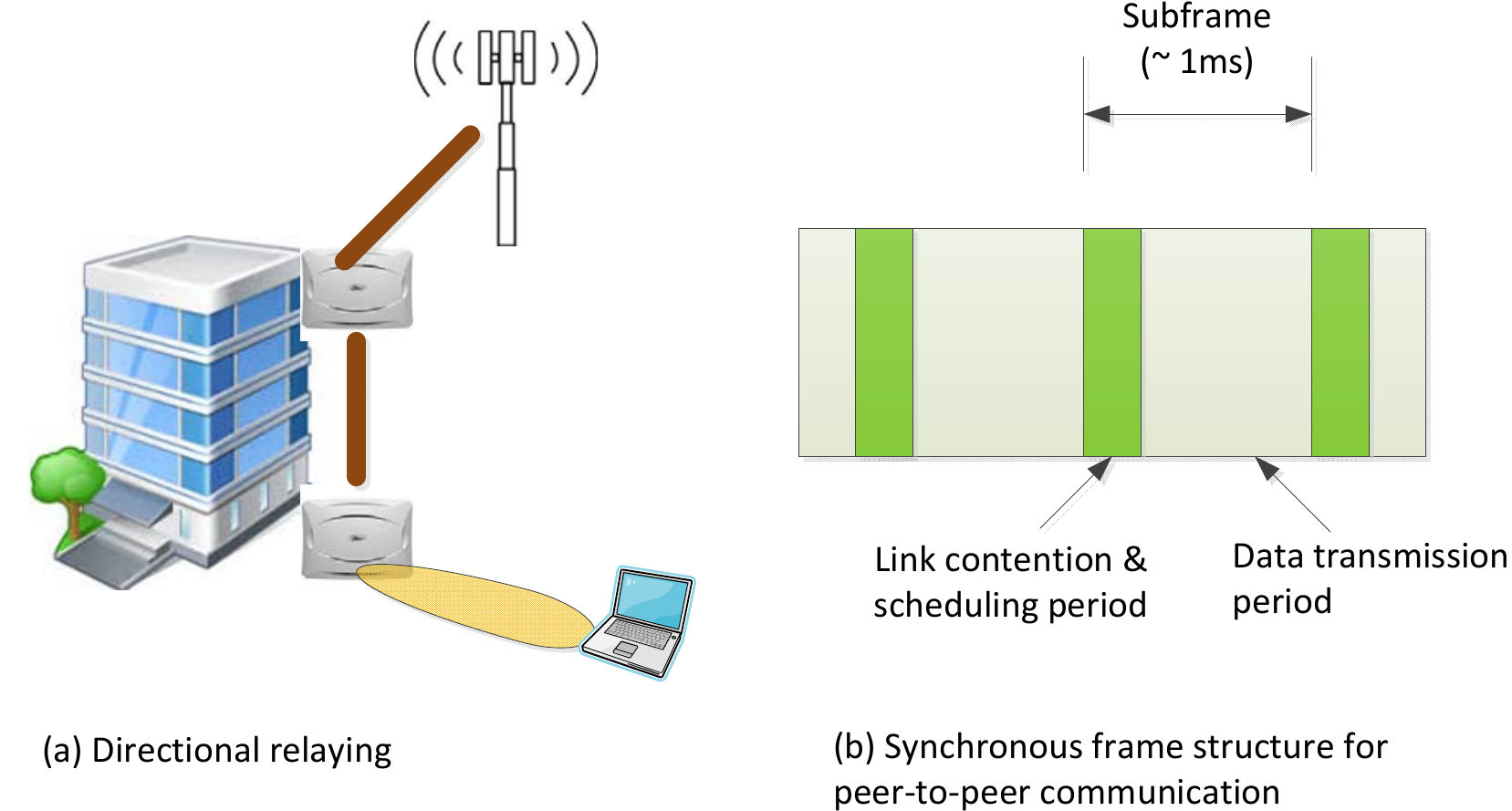}
\end{center}
\caption{\textbf{Directional mmW relaying:} (a) Multihop directional
relaying can provide wireless backhaul and
extend coverage of mmW signals in the presence of clutter and shadowing.
(b) A synchronous peer-to-peer frame structure along the lines of
\cite{FlashLinq:10-allerton} can enable
fast coordination and resource allocation across
relays, base stations and mobiles with dynamic duplexing. }
\label{fig:relaySys}
\end{figure}

Another key design issue for mmW cellular systems is support
for repeaters / relays -- a feature that can be particularly valuable
due to the need for range extension.
In current cellular systems,
relaying has been primarily used both for coverage extension
and, to a lesser extent, capacity expansion when backhaul is not available
\cite{SoLiang:05,SchoenenZW:08,BouSalehRRH:09}.
Although significant research went into enabling relaying in 3GPP
 LTE-Advanced \cite{peters2009relay}, the projected gains
have been modest.  In dense interference-limited environments,
the loss in degrees of freedom with half-duplex constraints and multiple transmissions
is typically not worth the increase in received power from shorter range.

With regards to relaying, mmW networks may be fundamentally different.
As discussed above, one of the most greatest challenges for mmW systems
is that mobiles may be in outage to the closest cell, dramatically reducing
the cell edge rate.
In these cases, relaying may be necessary to selectively
extend coverage to certain users and provide
a more uniform quality of service throughout the network.
Furthermore, given the inability of mmW signals to penetrate indoors,
relaying would also be essential to provide seamless indoor-outdoor coverage
and coverage in and around vehicles, airplanes, etc.
Relaying may also be valuable for backhaul to picocells when
fiber connectivity is not available
\cite{EricssonBackhaul:13,NGNM-Backhaul:12-short,ECC-Backhaul:12}.
Depending on the cell locations,
some of these mmW links may be in the clutter and require NLOS connectivity
similar to the access links -- See, for example, Fig.~\ref{fig:relaySys}(a).

In order to obtain the full advantages of relaying, cellular systems may need
to be significantly re-designed.
Cellular systems have traditionally followed a basic paradigm dividing
networks into distinct base stations and mobiles,
with relays typically being added as an afterthought.
However, given the central role that relaying may play in the mmW range for both the access link and for backhaul,
it may be worth investigating new \emph{peer-to-peer} topologies,
such as Qualcomm's FlashLinQ system \cite{FlashLinq:10-allerton},
where there is less centralized scheduling and where
frequency band and time slots
are not statically pre-allocated to traffic in any one direction.
As shown in Fig.~\ref{fig:relaySys}(b), one may consider symmetric frame structures
that are common in the uplink and downlink.  The directions of the links would not
necessarily need to be synchronized across the network, and a periodic
contention period can be used to reassign the directions of the links as necessary.
Such a design would be a significant departure from the uplink-downlink
in current LTE systems, but would enable much greater flexibility for multihop
networks and integrated systems for both access and backhaul.

\subsection{An End to Interference?}
As mentioned above, current cellular networks in dense urban deployments
are overwhelmingly interference-limited.  At a high level, mitigating this interference
can be seen as the driving motiviation behind many of the advanced technologies
introduced into cellular systems in the last decade.  These techniques
include coordinated
multipoint, inter-cellular interference coordination and more forward-looking concepts
such as interference alignment.

One of the striking conclusions of the above analysis is that many of these techniques
may have much more limited gains in the mmW space.  As we saw, for many mobiles,
thermal noise is significantly larger than interference.  That is, in mmW systems
with appropriate beamforming, links become directionally isolated and inter-cellular
interference is greatly reduced.  This fact implies that point-to-point, rather
than network, technologies may play a much larger role in achieving capacity gains
in these systems.

\subsection{Exploiting Channel Sparsity and Compressed Sensing} \label{sec:compSens}
As described in Section~\ref{sec:challenges}, one possible challenge in mmW
system is the high Doppler.  In general, Doppler spread is a function of the total
angular dispersion, carrier frequency and mobile velocity \cite{Rappaport:02}.
Thus, due to the high carrier frequencies and significant local scattering,
one might initially think that the total
Doppler spread in mmW systems will be high and potentially difficult to track.

However, the measurements reviewed in Section~\ref{sec:chanMeas}
revealed that signals generally arrive on a small number
of path clusters, each with relatively small angular spread.
Directional antennas will further reduce the multipath angular spread
~\cite{durgin2000theory}.
This property implies that the individually resolvable multipath components
will vary very slowly -- a fact confirmed directly in our
experiments in \cite{rappaportmillimeter}.  This is good news.

To understand how to exploit these slow variations for tracking the channel,
first observe that
the narrowband channel response at any particular frequency could be described as
\beq\label{eq:htsparse}
    h(t) = \sum_{k=1}^K g_k(t)e^{2\pi i f_{d}\cos(\theta_k)t},
\eeq
where $K$ is the number of clusters, $f_d$ is the
maximum Doppler shift, $\theta_k$ is the central angle of arrival
of the cluster and $g_k(t)$ is the time-varying gain of the channel related to the
angular spread \emph{within} the cluster.
Since the angular spread within each cluster is small,
the cluster gains $g_k(t)$ will generally be slowly
varying even though the aggregate channel $h(t)$ may have much higher variations.
Moreover, the angles of arrival $\theta_k$ are also typically slowly varying
since they are a result of the large scale scattering environment and do not change
with small scale mobility.  This fact suggests that even though $h(t)$ may change
rapidly, the parametrization \eqref{eq:htsparse} may enable more accurate tracking,
particularly since the number of clusters, $K$, tends to be small ($K$ is typically
1 to 5 in our measurements).

The parametrization \eqref{eq:htsparse} is fundamentally \emph{nonlinear}
and analogous to the types of models used in finite rate of innovation models
\cite{VetterliMB:02} and compressed sensing-based channel estimation
and channel sounding \cite{TaubockHlaw:08,HauptBRN:10,BarbotinHRV:11,tamir2012analog}.
The extension
of these methods to very wideband systems with large numbers of antennas may therefore
have significant value.

\subsection{Heterogeneous Networking Issues} \label{sec:hetNet}

As described in Section~\ref{sec:network},
mmW systems cannot be deployed in a standalone manner.
To provide uniform, reliable coverage, fallback to cellular
systems in conventional UHF or microwave frequencies will be necessary.
While support for heterogeneous networks has been a key design goal
in recent cellular standards,  mmW systems will push the need for support
for heterogeneous networks in several new directions.

Most importantly, the heterogeneous network support in mmW will require
cell selections and path switching at much faster rates than current
cellular systems.
Due to their vulnerability to shadowing,
mmW signals to any one cell will be inherently unreliable and can rapidly
change with small motions of the users or the user's environment.
One avenue to explore is the use \emph{carrier aggregation} techniques \cite{3GPP36.300,YuanZWY:10} where mobiles can connect to multiple
base stations simultaneously.  Carrier aggregation was introduced in release 10 of 3GPP LTE-Advanced primarily to increase peak throughputs.
For mmW systems, carrier aggregation could provide macro-diversity,
but would require support for path switching and scheduling in the network.

A second issue in the evolution of HetNets for mmW will be multi-operator
support.  Indoor cells and cells mounted on private buildings, may be better operated
by a third-party who would then provide roaming support for carriers from multiple
subscribers.  While roaming is commonly used in current networks, the time scales
for mmW roaming would be much faster.  In addition, with carrier aggregation, it may
be desirable for a mobile to be connected to cells from different operators
simultaneously.

Further complicating matters is the fact that, given the large amount of spectrum,
a single operator may not be able to fully utilize the bandwidth.
Thus, the model where a single operator has exclusive rights to a bandwidth
may not lead to the most efficient use of the spectrum.  However, support for
multiple operators sharing spectrum will need much more sophisticated inter-cell
interference coordination mechanisms, especially with directionality.
Future clearing houses will provide such measurement and management
for multiple carriers and their users.

\section{Conclusions}

Millimeter systems offer tremendous potential with orders
of magnitude greater spectrum and further gains from high-dimensional
antenna arrays.
To assess the feasibility of mmW systems, we have presented some initial
propagation measurements in NYC -- a challenging environment, but representative
of likely initial deployments.  Our measurements and capacity analysis
have revealed several surprising features:  Through reflections and scattering,
mmW signals are potentially viable
at distances of 100~m to 200~m, even in completely NLOS settings.
Moreover, with modest assumptions on beamforming,
our capacity analysis has indicated that mmW systems can offer at least
an order of magnitude in capacity over over current state-of-the-art LTE
systems, at least for outdoor coverage.

Potential mmW cellular systems may need to
be significantly re-designed relative to
current 4G systems to obtain the full potential
of mmW bands.  In particular,
the heavy reliance on directional transmissions and beamforming
will necessitate reconsideration of many basic procedures such as
cell search, synchronization, random access and intermittent communication.
Multiple access and channelization also become tied to front-end requirements,
particularly with regard to analog beamforming and A/D conversion.

In addition, directional isolation between links suggests that interference
mitigation, which has been dominant driver for new cellular technologies
in the last decade, may have a less significant impact in mmW.  On the other hand,
technologies such as carrier aggregation and multihop relaying that have
had only modest benefits in current cellular networks may play a very prominent
role in the mmW space.
These design issues -- though stemming from carrier frequency -- span all
the layers
of communication stack and will present a challenging, but exciting, set of
research problems that can ultimately revolutionize cellular communication.

\section*{Acknowledgements}
The authors would like to deeply thank several students and colleagues
for providing the propagation data
\cite{Rappaport:12-28G,Rappaport:28NYCPenetrationLoss,Samimi:AoAD,rappaportmillimeter,Sun-Beam:13,Nie72G-PIMRC:13,Rappaport:13-BBmmW}
and capacity analysis~\cite{AkLiuRanRapEr:13-arxiv}
that made this research possible:
Mutafa Riza Akdeniz, Yaniv Azar, Felix Gutierrez, DuckDong Hwang,
Yuanpeng Liu, Rimma Mayzus, George MacCartney, Shuai Nie,
Mathew K. Samimi, Jocelyn K. Schulz,  Shu Sun, Kevin Wang,
George N. Wong and  Hang Zhao.
This work also benefitted significantly from discussions with our
industrial affiliate partners in NYU WIRELESS, including Samsung, Intel, NSN, Qualcomm
and InterDigital.
Our work also benefitted from discussions from several researchers
including Jeffrey Andrews, Giuseppe Caire, Mung Chiang, Robert Heath, Upamanyu Madhow,
and Wei Yu, as well as the anonymous comments from the reviewers.

\bibliographystyle{IEEEtran}
\bibliography{bibl}

\newcommand{\SortNoop}[1]{}
\begin{thebibliography}{100}
\providecommand{\url}[1]{#1}
\csname url@samestyle\endcsname
\providecommand{\newblock}{\relax}
\providecommand{\bibinfo}[2]{#2}
\providecommand{\BIBentrySTDinterwordspacing}{\spaceskip=0pt\relax}
\providecommand{\BIBentryALTinterwordstretchfactor}{4}
\providecommand{\BIBentryALTinterwordspacing}{\spaceskip=\fontdimen2\font plus
\BIBentryALTinterwordstretchfactor\fontdimen3\font minus
  \fontdimen4\font\relax}
\providecommand{\BIBforeignlanguage}[2]{{%
\expandafter\ifx\csname l@#1\endcsname\relax
\typeout{** WARNING: IEEEtran.bst: No hyphenation pattern has been}%
\typeout{** loaded for the language `#1'. Using the pattern for}%
\typeout{** the default language instead.}%
\else
\language=\csname l@#1\endcsname
\fi
#2}}
\providecommand{\BIBdecl}{\relax}
\BIBdecl

\bibitem{CiscoVNI:latest}
Cisco, ``{Cisco Visual Network Index}: Global mobile traffic forecast update,''
  2013.

\bibitem{EricssonMDT:latest}
Ericsson, ``Traffic and market data report,'' 2011.

\bibitem{UMTSForecast}
{UMTS Forum}, ``Mobile traffic forecasts: 2010-2020 report,'' in \emph{UMTS
  Forum Report}, vol.~44, 2011.

\bibitem{Ericsson-50Billion}
{Ericsson}, ``More than 50 billion connected devices,'' Ericsson white paper,
  [online] http://www.ericsson.com/res/docs/whitepapers/wp-50-billions.pdf,
  2011.

\bibitem{NSN-B4G:12}
P.~E. Mogensen, K.~Pajukoski, B.~Raaf, E.~Tiirola, L.~Eva, I.~Z. Kovacs,
  G.~Berardinelli, L.~G.~U. Garcia, L.~Hu, and A.~F. Cattoni, ``{B4G} local
  area: high level requirements and system design,'' in \emph{Proc.\ IEEE
  Globecom Workshop}, Dec. 2012.

\bibitem{Ted:60Gstate11}
T.~S. Rappaport, J.~N. Murdock, and F.~Gutierrez, ``{State of the art in 60-GHz
  integrated circuits and systems for wireless communications},''
  \emph{Proceedings of the IEEE}, vol.~99, no.~8, pp. 1390 -- 1436, August
  2011.

\bibitem{KhanPi:11}
F.~Khan and Z.~Pi, ``{Millimeter-wave {M}obile {B}roadband ({MMB}):
  {U}nleashing 3-300GHz Spectrum},'' in \emph{Proc.\ IEEE Sarnoff Symposium},
  Mar. 2011.

\bibitem{KhanPi:11-CommMag}
------, ``{An introduction to millimeter-wave mobile broadband systems},''
  \emph{IEEE Comm. Mag.}, vol.~49, no.~6, pp. 101 -- 107, Jun. 2011.

\bibitem{PietBRPC:12}
P.~Pietraski, D.~Britz, A.~Roy, R.~Pragada, and G.~Charlton, ``Millimeter wave
  and terahertz communications: Feasibility and challenges,'' \emph{ZTE
  Communications}, vol.~10, no.~4, pp. 3--12, Dec. 2012.

\bibitem{Doan:04}
C.~Doan, S.~Emami, D.~Sobel, A.~Niknejad, and R.~Brodersen, ``{Design
  considerations for 60 GHz CMOS radios},'' \emph{IEEE Comm. Mag.}, vol.~42,
  no.~12, pp. 132 -- 140, 2004.

\bibitem{Doan:05}
C.~Doan, S.~Emami, A.~Niknejad, and R.~Brodersen, ``Millimeter-wave {CMOS}
  design,'' \emph{IEEE J. Solid-State Circuts}, vol.~40, no.~1, pp. 144--155,
  2005.

\bibitem{ZhaLiu:09}
Y.-P. Zhang and D.~Liu, ``Antenna-on-{C}hip and {A}ntenna-in-{P}ackage
  solutions to highly integrated millimeter-wave devices for wireless
  communications,'' \emph{IEEE Trans.\ Antennas and Propagation}, vol.~57,
  no.~10, pp. 2830--2841, 2009.

\bibitem{gutierrez2009chip}
F.~Gutierrez, S.~Agarwal, K.~Parrish, and T.~S. Rappaport, ``On-chip integrated
  antenna structures in {CMOS} for 60 {GHz WPAN} systems,'' \emph{IEEE J. Sel.
  Areas Comm.}, vol.~27, no.~8, pp. 1367--1378, 2009.

\bibitem{Nsenga:10}
J.~Nsenga, A.~Bourdoux, and F.~Horlin, ``Mixed analog/digital beamforming for
  60 {GHz MIMO} frequency selective channels,'' in \emph{Proc.\ IEEE ICC},
  2010.

\bibitem{Rajagopal:mmWMobile}
S.~Rajagopal, S.~Abu-Surra, Z.~Pi, and F.~Khan, ``Antenna array design for
  multi-gbps mmwave mobile broadband communication,'' in \emph{Proc.\ IEEE
  Globecom}, 2011.

\bibitem{Huang:2008:MWA:1524107}
K.-C. Huang and D.~J. Edwards, \emph{Millimetre Wave Antennas for Gigabit
  Wireless Communications: A Practical Guide to Design and Analysis in a System
  Context}.\hskip 1em plus 0.5em minus 0.4em\relax Wiley Publishing, 2008.

\bibitem{Rusek:13}
F.~Rusek, D.~Persson, B.~K. Lau, E.~Larsson, T.~Marzetta, O.~Edfors, and
  F.~Tufvesson, ``Scaling up {MIMO}: Opportunities and challenges with very
  large arrays,'' \emph{IEEE Signal Process. Mag.}, vol.~30, no.~1, pp. 40--60,
  2013.

\bibitem{hur2013millimeter}
S.~Hur, T.~Kim, D.~J. Love, J.~V. Krogmeier, T.~A. Thomas, and A.~Ghosh,
  ``Millimeter wave beamforming for wireless backhaul and access in small cell
  networks,'' \emph{arXiv preprint arXiv:1306.6659}, 2013.

\bibitem{Samsung5G:13}
Samsung, ``Samsung announces world’s first 5g mmwave mobile technology,''
  Press release at http://global.samsungtomorrow.com/?p=24093, May 2013.

\bibitem{fettweis2005wigwam}
G.~Fettweis and R.~Irmer, ``{WIGWAM: S}ystem concept development for 1 {G}bit/s
  air interface,'' in \emph{Proc.\ Wireless World Research Forum}, 2005.

\bibitem{laskar2007next}
J.~Laskar, S.~Pinel, D.~Dawn, S.~Sarkar, B.~Perumana, and P.~Sen, ``The next
  wireless wave is a millimeter wave,'' \emph{Microwave Journal}, vol.~50,
  no.~8, p.~22, 2007.

\bibitem{AkoumAyaHeath:12}
S.~Akoum, O.~E. Ayach, and R.~W. Heath, ``{Coverage and capacity in mmWave MIMO
  systems},'' in \emph{Proc. of Asilomar Conf. on Signals, Syst. \& Computers},
  Pacific Grove, CA, Nov. 2012.

\bibitem{ZhangMadhow1}
H.~Zhang, S.~Venkateswaran, and U.~Madhow, ``{Channel modeling and {MIMO}
  capacity for outdoor millimeter wave links},'' in \emph{Proc. IEEE WCNC},
  April 2010.

\bibitem{ZhangMadhow2}
E.~Torkildson, H.~Zhang, and U.~Madhow, ``{Channel modeling for millimeter wave
  MIMO},'' in \emph{Proc.\ Information Theory and Applications Workshop (ITA)},
  Feb. 5 2010.

\bibitem{ZhangMadhow3}
H.~Zhang and U.~Madhow, ``{Statistical modeling of fading and diversity for
  outdoor 60 GHz channels},'' in \emph{International Workshop on mmWave
  Communications: from Circuits to Networks (mmCom10)}, September 2010.

\bibitem{rappaportmillimeter}
T.~S. Rappaport, S.~Sun, R.~Mayzus, H.~Zhao, Y.~Azar, K.~Wang, G.~N. Wong,
  J.~K. Schulz, M.~Samimi, and F.~Gutierrez, ``{Millimeter Wave Mobile
  Communications for 5G Cellular: It Will Work!}'' \emph{IEEE Access}, vol.~1,
  pp. 335--349, May 2013.

\bibitem{BocHLMP:14}
F.~Boccardi, R.~W. Heath, A.~Lozano, T.~L. Marzetta, and P.~Popovski, ``Five
  disruptive technology directions for {5G},'' to appear in \emph{IEEE Comm.
  Magazine}, 2014.

\bibitem{ben2011millimeter}
E.~Ben-Dor, T.~S. Rappaport, Y.~Qiao, and S.~J. Lauffenburger,
  ``Millimeter-wave {60 GH}z outdoor and vehicle {AOA} propagation measurements
  using a broadband channel sounder,'' in \emph{Proc.\ IEEE Globecom}, 2011,
  pp. 1--6.

\bibitem{Rappaport:12-28G}
Y.~Azar, G.~N. Wong, K.~Wang, R.~Mayzus, J.~K. Schulz, H.~Zhao, F.~Gutierrez,
  D.~Hwang, and T.~S. Rappaport, ``28 {GHz} propagation measurements for
  outdoor cellular communications using steerable beam antennas in {N}ew {Y}ork
  {C}ity,'' in \emph{Proc.\ IEEE ICC}, 2013.

\bibitem{Rappaport:28NYCPenetrationLoss}
H.~Zhao, R.~Mayzus, S.~Sun, M.~Samimi, J.~K. Schulz, Y.~Azar, K.~Wang, G.~N.
  Wong, F.~Gutierrez, and T.~S. Rappaport, ``28 {GHz} millimeter wave cellular
  communication measurements for reflection and penetration loss in and around
  buildings in {N}ew {Y}ork {C}ity,'' in \emph{Proc. IEEE ICC}, 2013.

\bibitem{Samimi:AoAD}
M.~Samimi, K.~Wang, Y.~Azar, G.~N. Wong, R.~Mayzus, H.~Zhao, J.~K. Schulz,
  S.~Sun, F.~Gutierrez, and T.~S. Rappaport, ``28 {GHz} angle of arrival and
  angle of departure analysis for outdoor cellular communications using
  steerable beam antennas in {N}ew {Y}ork {C}ity,'' in \emph{Proc. IEEE VTC},
  2013.

\bibitem{Nie72G-PIMRC:13}
S.~Nie, G.~R. {MacCartney, Jr}., S.~Sun, and T.~S. Rappaport, ``72 {GH}z
  millimeter wave indoor measurements for wireless and backhaul
  communications,'' in \emph{Proc. IEEE PIMRC}, Sep. 2013.

\bibitem{Rappaport:13-BBmmW}
T.~S. Rappaport, F.~Gutierrez, E.~Ben-Dor, J.~N. Murdock, Y.~Qiao, and J.~I.
  Tamir, ``Broadband millimeter-wave propagation measurements and models using
  adaptive-beam antennas for outdoor urban cellular communications,''
  \emph{IEEE Trans.\ Antennas and Propagation}, vol.~61, no.~4, pp. 1850--1859,
  2013.

\bibitem{AkLiuRanRapEr:13-arxiv}
M.~R. Akdeniz, Y.~Liu, M.~K. Samimi, S.~Sun, S.~Rangan, T.~S. Rappaport, and
  E.~Erkip, ``Millimeter wave channel modeling and cellular capacity
  evaluation,'' http://arxiv.org/abs/1312.4921, Dec. 2013.

\bibitem{Bose:27}
J.~Bose, \emph{Collected Physical Papers}.\hskip 1em plus 0.5em minus
  0.4em\relax New York, NY: Longmans, Green and Co., 1927.

\bibitem{Roddy:06}
D.~Roddy, \emph{Satellite Communications}, 4th~ed.\hskip 1em plus 0.5em minus
  0.4em\relax McGraw Hill, 2006.

\bibitem{EricssonBackhaul:13}
J.~Hansry, J.~Edstam, B.-E. Olsson, and C.~Larsson, ``Non-line-of-sight
  microwave backhaul for small cells,'' \emph{Ericsson Review}, Feb. 2013.

\bibitem{NGNM-Backhaul:12-short}
{NGMN Alliance}, ``Small cell backhaul requirements,'' Whitepaper available at
  http://www.ngmn.org/uploads/media, Jun. 2012.

\bibitem{ECC-Backhaul:12}
{Electronic Communications Committee (ECC)}, ``Fixed service in europe –
  current use and future trends,'' Available at
  http://www.erodocdb.dk/Docs/doc98/official/pdf/ECCRep173.PDF, 2012.

\bibitem{PerCPY:10}
E.~Perahia, C.~Cordeiro, M.~Park, and L.~Yang, ``{IEEE 802.11ad: Defining the
  next generation multi-Gbps Wi-Fi},'' in \emph{Proc.\ IEEE Cons.\ Comm.\ \&
  Network.\ Conf.}, Jan. 2010.

\bibitem{VauNic:10}
S.~J. Vaughan-Nichols, ``{Gigabit Wi-Fi is on its way},'' \emph{IEEE Computer},
  Nov. 2010.

\bibitem{Daniels:10}
R.~Daniels, J.~Murdock, T.~S. Rappaport, and R.~Heath, ``{60 GHz Wireless: Up
  Close and Personal},'' \emph{IEEE Microwave Magazine}, vol.~11, no.~7, pp. 44
  -- 50, December 2010.

\bibitem{Baykas-WPAN:11}
T.~Baykas, C.-S. Sum, Z.~Lan, J.~Wang, M.~A. Rahman, H.~Harada, and S.~Kato,
  ``{IEEE 802.15.3c: The first IEEE wireless standard for data rates over 1
  Gb/s},'' \emph{IEEE Comm. Mag.}, Jul. 2011.

\bibitem{Ortiz:08}
S.~Ortiz, ``The wireless industry begins to embrace femtocells,'' \emph{IEEE
  Computer}, vol.~41, no.~7, pp. 14--17, Jul. 2008.

\bibitem{ChaAndG:08}
V.~Chandrasekhar, J.~G. Andrews, and A.~Gatherer, ``Femtocell networks: A
  survey,'' \emph{IEEE Comm. Mag.}, vol.~46, no.~9, pp. 59--67, Sep. 2009.

\bibitem{YehTLK:08}
S.-P. Yeh, S.~Talwar, S.-C. Lee, and H.~Kim, ``{WiMAX} femtocells: A
  perspective on network architecture, capacity, and coverage,'' \emph{IEEE
  Comm. Mag.}, vol.~46, no.~10, pp. 58--65, Oct. 2008.

\bibitem{FemtoForum:10}
{Femto Forum}, ``Interference management in {OFDMA} femtocells,'' Whitepaper
  available at www.femtoforum.org, Mar. 2010.

\bibitem{AndrewsCDRC:12}
J.~G. Andrews, H.~Claussen, M.~Dohler, S.~Rangan, and M.~C. Reed, ``Femtocells:
  Past, present, and future,'' \emph{IEEE J. Sel. Areas Comm.}, vol.~30, no.~3,
  Apr. 2012.

\bibitem{Qualcomm-SmallCell}
{Qualcomm}, ``The 1000x challenge: More small cells -- taking hetnets to the
  next level,'' [online]
  http://www.qualcomm.com/solutions/wireless-networks/technologies/1000x-data/small-cells,
  2013.

\bibitem{financial_picos}
H.~Claussen, L.~T.~W. Ho, and L.~Samuel, ``Financial analysis of a
  pico-cellular home network deployment,'' in \emph{Proc.\ IEEE ICC}, Jun.
  2007, pp. 5604--5609.

\bibitem{senza_backhaul}
{Senza Fili Consulting}, ``Crucial economics for mobile data backhaul,''
  Whitepaper available at
  http://cbnl.com/sites/all/files/userfiles/files/CB-002070-DC-LATEST.pdf,
  2011.

\bibitem{backhaul_blog_1}
D.~Webster, ``Solving the mobile backhaul bottleneck,''
  http://blogs.cisco.com/sp/solving\_the\_mobile\_backhaul\_bottleneck, Apr.
  2009.

\bibitem{backhaul_blog_2}
C.~Mathias, ``Fixing the cellular network: Backhaul is the key,''
  http://www.networkworld.com/community/print/35920, Dec. 2008.

\bibitem{Rappaport:02}
T.~S. Rappaport, \emph{Wireless Communications: Principles and Practice},
  2nd~ed.\hskip 1em plus 0.5em minus 0.4em\relax Upper Saddle River, NJ:
  Prentice Hall, 2002.

\bibitem{Sun-Beam:13}
S.~Sun and T.~S. Rappaport, ``Multi-beam antenna combining for 28 {GHz}
  cellular link improvement in urban environments,'' in \emph{Proc. IEEE
  Globecom}, Dec. 2013.

\bibitem{Allen:94}
K.~Allen \emph{et~al.}, \emph{{Building penetration loss measurements at 900
  MHz, 11.4 GHz, and 28.8 MHz}}, ser. NTIA report -- 94-306.\hskip 1em plus
  0.5em minus 0.4em\relax {Boulder, CO}: U.S. Dept. of Commerce, National
  Telecommunications and Information Administration, 1994.

\bibitem{Anderson04}
C.~R. Anderson and T.~S. Rappaport, ``{In-building wideband partition loss
  measurements at 2.5 and 60 GHz},'' \emph{IEEE Trans. Wireless Comm.}, vol.~3,
  no.~3, pp. 922 -- 928, May 2004.

\bibitem{Alejos:08}
A.~Alejos, M.~Sanchez, and I.~Cuinas, ``Measurement and analysis of propagation
  mechanisms at 40 {GHz}: Viability of site shielding forced by obstacles,''
  \emph{IEEE Trans.\ Vehicular Technology}, vol.~57, no.~6, pp. 3369--3380,
  2008.

\bibitem{LuSCP:12}
J.~S. Lu, D.~Steinbach, P.~Cabrol, and P.~Pietraski, ``Modeling human blockers
  in millimeter wave radio links,'' \emph{ZTE Communications}, vol.~10, no.~4,
  pp. 23--28, Dec. 2012.

\bibitem{ChiaGB:09}
S.~Chia, M.~Gasparroni, and P.~Brick, ``The next challenge for cellular
  networks: backhaul,'' \emph{IEEE Microwave Magazine}, vol.~10, no.~5, pp.
  54--66, May 2009.

\bibitem{ChoEtAl_LowPowerAD:94}
T.~Cho, D.~Cline, C.~Conroy, and P.~Gray, ``{Design considerations for
  low-power, high-speed CMOS analog/digital converters},'' in \emph{Proc.\ IEEE
  Symp.\ Low Power Electronics}, Oct. 1994.

\bibitem{murdock2014consumption}
J.~Murdock and T.~S. Rappaport, ``Consumption factor and power-efficiency
  factor: A theory for evaluating the energy efficiency of cascaded
  communication systems,'' \emph{IEEE J. Sel. Areas Comm.}, Dec. 2014.

\bibitem{ChenWHL:11}
C.-Y. Chen, J.~Wu, J.-J. Hung, T.~Li, W.~Liu, and W.-T. Shih, ``{A 12-Bit 3
  GS/s Pipeline ADC With 0.4$^2$ mm and 500 mW in 40 nm Digital CMOS},''
  \emph{IEEE J. Solid-State Circuts}, vol.~47, no.~4, pp. 1013--1021, apr 2011.

\bibitem{ParkEtAl:11}
H.-L. Park, Y.-G. Kwon, M.-H. Choi, Y.~Kim, S.-H. Lee, Y.-D. Jeon, and J.-K.
  Kwon, ``{A 6b 1.2 GS/s 47.8 mW 0.17 mm$^2$ 65 nm CMOS ADC for High-Rate WPAN
  Systems},'' \emph{J.\ Semiconductor Sci. \& Tech.}, vol.~11, no.~2, Jun.
  2011.

\bibitem{abouelseoud2013system}
M.~Abouelseoud and G.~Charlton, ``System level performance of millimeter-wave
  access link for outdoor coverage,'' in \emph{Proc.\ IEEE WCNC}, 2013, pp.
  4146--4151.

\bibitem{Ghosh-mmw:2013}
A.~Ghosh, ``Can mmwave wireless technology meet the future capacity crunch?''
  Presentation at IEEE ICC 2013, available at
  http://www.ieee-icc.org/2013/Mmwave\_Spring\_ICC2013\_Ghosh.pdf, Budapest,
  Hungary, Jun. 2013.

\bibitem{QualcommHetNet:11}
Qualcomm, ``{LTE} {A}dvanced: {H}eterogeneous networks,'' Whitepaper available
  online at http://www.qualcomm.com/media/documents, Jan. 2011.

\bibitem{QCOM-HetNetSurvey:11}
A.~Damnjanovic, J.~Montojo, Y.~Wei, T.~Ji, T.~Luo, M.~Vajapeyam, T.~Yoo,
  O.~Song, and D.~Malladi, ``A survey on {3GPP} heterogeneous networks,''
  \emph{IEEE Wireless Communications}, vol.~18, no.~3, pp. 10--21, 2011.

\bibitem{Qualcomm-NSC}
{Qualcomm}, ``Neighborhood small cells for hyperdense deployments: Taking
  hetnets to the next level,'' [online]
  http://www.qualcomm.com/media/documents/files/qualcomm-research-neighborhood-small-cell-deployment-model.pdf,
  Feb. 2013.

\bibitem{FordKR:13-arxiv}
R.~Ford, C.~Kim, and S.~Rangan, ``Optimal user association in cellular networks
  with opportunistic third-party backhaul,'' arXiv preprint, Apr. 2013.

\bibitem{Zwick05}
T.~Zwick, T.~Beukema, and H.~Nam, ``Wideband channel sounder with measurements
  and model for the 60 {GHz} indoor radio channel,'' \emph{IEEE Trans.\
  Vehicular Technology}, vol.~54, no.~4, pp. 1266 -- 1277, July 2005.

\bibitem{Giannetti:99}
F.~Giannetti, M.~Luise, and R.~Reggiannini, ``{Mobile and personal
  communications in 60 GHz band: A survey},'' \emph{Wirelesss Personal
  Comunications}, vol.~10, pp. 207 -- 243, 1999.

\bibitem{Smulders}
P.~Smulders and A.~Wagemans, ``{Wideband indoor radio propagation measurements
  at 58 GHz},'' \emph{Electronics Letters}, vol.~28, no.~13, pp. 1270 --1272,
  June 1992.

\bibitem{Manabe}
T.~Manabe, Y.~Miura, and T.~Ihara, ``{Effects of antenna directivity and
  polarization on indoor multipath propagation characteristics at 60 GHz},''
  \emph{IEEE J. Sel. Areas Comm.}, vol.~14, no.~3, pp. 441 --448, April 1996.

\bibitem{ted2}
H.~Xu, V.~Kukshya, and T.~S. Rappaport, ``{Spatial and temporal characteristics
  of 60 GHz indoor channel},'' \emph{IEEE J. Sel. Areas Comm.}, vol.~20, no.~3,
  pp. 620 -- 630, April 2002.

\bibitem{ElrefShak:97}
A.~Elrefaie and M.~Shakouri, ``Propagation measurements at 28 {GH}z for
  coverage evaluation of {L}ocal {M}ultipoint {D}istribution {S}ervice,''
  \emph{Proc.\ Wireless Communications Conference}, pp. 12--17, Aug. 1997.

\bibitem{SeiArn:95}
S.~Seidel and H.~Arnold, ``Propagation measurements at 28 ghz to investigate
  the performance of local multipoint distribution service (lmds),''
  \emph{Proc.\ Wireless Communications Conference}, pp. 754--757, Nov. 1995.

\bibitem{Rappaport38:12}
T.~S. Rappaport, E.~Ben-Dor, J.~Murdock, and Y.~Qiao, ``38 {GH}z and 60 {GH}z
  angle-dependent propagation for cellular and peer-to-peer wireless
  communications,'' \emph{Proc.\ IEEE ICC}, pp. 4568--4573, Jun. 2012.

\bibitem{ted:rww12}
T.~S. Rappaport, E.~Ben-Dor, J.~Murdock, Y.~Qiao, and J.~Tamir, ``{Cellular
  Broadband Millimeter Wave Propagation and Angle of Arrival for Adaptive Beam
  Steering Systems},'' in \emph{Proc. IEEE RWS (Invited)}, Jan. 2012.

\bibitem{ted:wcnc12}
J.~Murdock, E.~Ben-Dor, Y.~Qiao, J.~Tamir, and T.~S. Rappaport, ``{A 38 GHz
  Cellular Outage Study for an Urban Outdoor Campus Environment},'' in
  \emph{Proc. IEEE WCNC}, April 2012.

\bibitem{Rappaport-72GHz:13}
G.~R. {MacCartney, Jr.}, J.~Zhang, S.~Nie, and T.~S. Rappaport, ``Path loss
  models for {5G} millimeter wave propagation channels in urban microcells,''
  in \emph{Proc. IEEE Globecom}, Dec. 2013.

\bibitem{wells:09}
J.~A. Wells, ``Faster than fiber: The future of multi-{G}b/s wireless,''
  \emph{IEEE Microwave Magazine}, vol.~10, no.~3, pp. 104--112, May 2009.

\bibitem{3GPP36.814}
3GPP, ``{Further advancements for {E-UTRA} physical layer aspects},'' TR 36.814
  (release 9), 2010.

\bibitem{Z3}
S.~Pinel, S.~Sarkar, P.~Sen, B.~Perumana, D.~Yeh, D.~Dawn, and J.~Laskar, ``{A
  90nm {CMOS} 60 {GH}z Radio},'' in \emph{Proc.\ IEEE International Solid-State
  Circuits Conference}, 2008.

\bibitem{Z4}
C.~Marcu, D.~Chowdhury, C.~Thakkar, J.-D. Park, L.-K. Kong, M.~Tabesh, Y.~Wang,
  B.~Afshar, A.~Gupta, A.~Arbabian, , S.~Gambini, R.~Zamani, E.~Alon, and
  A.~Niknejad, ``A 90 nm {CMOS} low-power 60 {GH}z transceiver with integrated
  baseband circuitry,'' \emph{IEEE J. Solid-State Circuts}, vol.~44, pp.
  3434--3447, 2009.

\bibitem{Heath:partialBF}
A.~Alkhateeb, O.~E. Ayach, G.~Leus, and J.~Robert W.~Heath, ``Hybrid
  analog-digital beamforming design for millimeter wave cellular systems with
  partial channel knowledge,'' in \emph{Proc.\ Information Theory and
  Applications Workshop (ITA)}, Feb. 2013.

\bibitem{Lozano:07}
A.~Lozano, ``Long-term transmit beamforming for wireless multicasting,'' in
  \emph{Proc.\ ICASSP}, vol.~3, 2007, pp. III--417--III--420.

\bibitem{LimmWBooster:13}
Q.~Li, H.~Niu, G.~Wu, and R.~Q. Hu, ``Anchor-booster based heterogeneous
  networks with mmwave capable booster cells,'' in \emph{Proc.\ IEEE Globecom
  Workshop}, Dec. 2013.

\bibitem{bontu2009drx}
C.~Bontu and E.~Illidge, ``{DRX} mechanism for power saving in {LTE},''
  \emph{IEEE Communications Magazine}, vol.~47, no.~6, pp. 48--55, 2009.

\bibitem{ParkZim:02}
D.~Parker and D.~Z. Zimmermann, ``{Phased arrays-part I: Theory and
  architecture},'' \emph{IEEE Trans.\ Microw.\ Theory Tech.}, vol.~50, no.~3,
  pp. 678--687, Mar. 2002.

\bibitem{KohReb:07}
K.-J. Koh and G.~M. Rebeiz, ``{0.13- m CMOS phase shifters for X-, Ku- and
  K-band phased arrays},'' \emph{IEEE J. Solid-State Circuts}, vol.~42, no.~11,
  pp. 2535--2546, Nov. 2007.

\bibitem{KohReb:09}
------, ``{A Millimeter-Wave (40–45 GHz) 16-Element Phased-Array Transmitter
  in 0.18-m SiGe BiCMOS Technology},'' \emph{IEEE J. Solid-State Circuts},
  vol.~44, no.~5, pp. 1498--1509, May 2009.

\bibitem{Crane-Patent:88}
P.~E. Crane, ``{Phased array scanning system},'' United States Patent
  4,731,614, filed Aug 11, 1986, issued Mar.\ 15, 1988.

\bibitem{RamBaRe:98}
S.~Raman, N.~S. Barker, and G.~M. Rebeiz, ``{A W-band dielectriclens-based
  integrated monopulse radar receive},'' \emph{IEEE Trans.\ Microw.\ Theory
  Tech.}, vol.~46, no.~12, pp. 2308--2316, Dec. 1998.

\bibitem{GuanHaHa:04}
X.~Guan, H.~Hashemi, and A.~Hajimiri, ``{A fully integrated 24-GHz
  eight-element phased-array receiver in silicon},'' \emph{IEEE J. Solid-State
  Circuts}, vol.~39, no.~12, pp. 2311--2320, Dec. 2004.

\bibitem{fettweis2011entering}
G.~Fettweis, F.~Guderian, and S.~Krone, ``Entering the path towards terabit/s
  wireless links,'' in \emph{Proc.\ IEEE Design, Automation \& Test in Europe},
  2011, pp. 1--6.

\bibitem{Madhow:ADC}
H.~Zhang, S.~Venkateswaran, and U.~Madhow, ``Analog multitone with interference
  suppression: Relieving the {ADC} bottleneck for wideband 60 ghz systems,'' in
  \emph{Proc.\ IEEE Globecom}, Nov. 2012.

\bibitem{FlashLinq:10-allerton}
S.~Tavildar, S.~Shakkottai, T.~Richardson, J.~Li, R.~Laroia, and A.~Jovicic,
  ``Flash{L}in{Q}: A synchronous distributed scheduler for peer-to-peer ad hoc
  networks,'' in \emph{Proc.\ Allerton Conf.\ Comm.\ Control \& Comp.},
  Allerton, IL, Oct. 2010.

\bibitem{SoLiang:05}
A.~So and B.~Liang, ``{Effect of Relaying on Capacity Improvement in Wireless
  Local Area Networks},'' \emph{Proc.\ IEEE WCNC}, vol.~3, pp. 1539--1544,
  March 2005.

\bibitem{SchoenenZW:08}
R.~Schoenen, W.~Zirwas, and B.~H. Walke, ``{Capacity and Coverage Analysis of a
  3GPP-LTE Multihop Deployment Scenario},'' \emph{Proc.\ IEEE ICC}, pp. 31--36,
  May 2008.

\bibitem{BouSalehRRH:09}
A.~B. Saleh, S.~Redana, B.~Raaf, and J.~Hamalainen, ``{Comparison of Relay and
  Pico eNB Deployments in LTE-Advanced},'' \emph{Proc. IEEE Vehicular
  Technology Conference (VTC)}, September 2009.

\bibitem{peters2009relay}
S.~W. Peters, A.~Y. Panah, K.~T. Truong, and R.~W. Heath, ``Relay architectures
  for 3gpp lte-advanced,'' \emph{EURASIP Journal on Wireless Communications and
  Networking}, vol. 2009, p.~1, 2009.

\bibitem{durgin2000theory}
G.~D. Durgin and T.~S. Rappaport, ``Theory of multipath shape factors for
  small-scale fading wireless channels,'' \emph{IEEE Trans.\ on Antennas and
  Propagation}, vol.~48, no.~5, pp. 682--693, 2000.

\bibitem{VetterliMB:02}
M.~Vetterli, P.~Marziliano, and T.~Blu, ``Sampling signals with finite rate of
  innovation,'' \emph{IEEE Trans. Signal Process.}, vol.~50, no.~6, pp.
  1417--1428, Jun. 2002.

\bibitem{TaubockHlaw:08}
G.~Taubock and F.~Hlawatsch, ``A compressed sensing technique for {OFDM}
  channel estimation in mobile environments: Exploiting channel sparsity for
  reducing pilots,'' in \emph{Proc. IEEE Int. Conf. Acoust., Speech, and Signal
  Process.}, May 2008, pp. 2885 -- 2888.

\bibitem{HauptBRN:10}
J.~Haupt, W.~U. Bajwa, G.~Raz, and R.~Nowak, ``Toeplitz compressed sensing
  matrices with applications to sparse channel estimation,'' \emph{IEEE Trans.
  Inform. Theory}, vol.~56, no.~11, pp. 5862--5875, Nov. 2010.

\bibitem{BarbotinHRV:11}
Y.~Barbotin, A.~Hormati, S.~Rangan, and M.~Vetterli, ``Estimating sparse {MIMO}
  channels having common support,'' in \emph{Proc. IEEE Int. Conf. Acoust.,
  Speech, and Signal Process.}, May 2011, pp. 2920 -- 2923.

\bibitem{tamir2012analog}
J.~I. Tamir, T.~S. Rappaport, Y.~C. Eldar, and A.~Aziz, ``Analog compressed
  sensing for {RF} propagation channel sounding,'' in \emph{Proc.\ IEEE
  ICASSP}, 2012, pp. 5317--5320.

\bibitem{3GPP36.300}
3GPP, ``{Evolved Universal Terrestrial Radio Access (E-UTRA) and Evolved
  Universal Terrestrial Radio Access Network (E-UTRAN); Overall description;
  Stage 2},'' TS 36.300 (release 10), 2010.

\bibitem{YuanZWY:10}
G.~Yuan, X.~Zhang, W.~Wang, and Y.~Yang, ``Carrier aggregation for
  {LTE-A}dvanced mobile communication systems,'' \emph{IEEE Comm. Mag.},
  vol.~48, no.~2, pp. 88--93, 2010.

\end{thebibliography}

\end{document}